\documentclass[preprint,11pt,floats,epsf,superscriptaddress,nofootinbib,aps,prd]{revtex4-2}
\usepackage{graphicx}
\usepackage{dcolumn} 
\usepackage{array}
\usepackage{bm}
\usepackage{bbm}
\usepackage{amssymb}
\usepackage{amsmath}
\usepackage{epsfig}    
\usepackage{color}
\usepackage{xcolor}
\usepackage{slashed}
\usepackage{multirow}
\usepackage[normalem]{ulem}
\usepackage{physics}
\usepackage{tikz-feynman}
\usepackage{xspace}
\usepackage{comment}
\numberwithin{equation}{section}
\usepackage[colorlinks=true,
            linkcolor=red,
            filecolor=magenta,
            citecolor=blue,
            urlcolor=magenta
            ]{hyperref} % For creating hyperlinks in cross references
\let\oldfootnote\footnote
\renewcommand{\footnote}[1]{%
    \begingroup%
    \linespread{1}%    % <- linespread for footnote: 1, 1.1, 1.2 etc
    \oldfootnote{#1}%
    \endgroup%
}
\newcommand{\ie}{{i.e.\xspace }}
\newcommand{\eg}{{e.g.\xspace }}

\newcommand{\gev}{\,\mathrm{GeV}}
\newcommand{\tev}{\,\mathrm{TeV}}
\newcommand{\citere}[1]{Ref.~\cite{#1}}
\newcommand{\citeres}[1]{Refs.~\cite{#1}}
\newcommand{\refse}[1]{Sec.~\ref{#1}}
\newcommand{\refta}[1]{Tab.~\ref{#1}}
\newcommand{\reffi}[1]{Fig.~\ref{#1}}
\newcommand{\reffis}[1]{Figs.~\ref{#1}}
\newcommand{\refeq}[1]{Eq.~(\ref{#1})}
\newcommand{\refeqs}[1]{Eqs.~(\ref{#1})}

\newcommand{\muAC}{{0.24}}
\newcommand{\dmuACp}{{0.09}}
\newcommand{\dmuACm}{{0.08}}
\newcommand{\sigAC}{{3.1}}

\newcommand{\massAC}{{95.4}}

\newcommand{\megm}{meGM\xspace}

\definecolor{Orange}{named}{orange}
\definecolor{Red}{named}{red}
\definecolor{Blue}{named}{blue}

\usepackage{cancel}
\allowdisplaybreaks

\begin{document}

\count\footins = 1000

%================================================================
% Front Matter
%================================================================
\preprint{DESY-25-159, IFT--UAM/CSIC-25-140}

\title{Interpretation
of LHC excesses at \boldmath{$95 \gev$ and $152 \gev$}\\[.5em] 
in an extended Georgi-Machacek model}

\author{Ting-Kuo Chen}
\email{tchen463@wisc.edu}
\affiliation{Department of Physics, University of Wisconsin-Madison, Madison, WI 53706, USA}

\author{Cheng-Wei Chiang}
\email{chengwei@phys.ntu.edu.tw}
\affiliation{Department of Physics and Center for Theoretical Physics, National Taiwan University, Taipei, Taiwan 10617, ROC}
\affiliation{Physics Division, National Center for Theoretical Sciences, Taipei, Taiwan 10617, ROC}

\author{Sven Heinemeyer}
\email{sven.heinemeyer@cern.ch}
\affiliation{Instituto de Física Teórica UAM-CSIC, Cantoblanco, 28049, Madrid, Spain}

\author{Georg Weiglein}
\email{georg.weiglein@desy.de}
\affiliation{Deutsches\,Elektronen-Synchrotron\,DESY,\,Notkestraße\,85,\,22607\,Hamburg,\,Germany}
\affiliation{II.~Institut für Theoretische Physik, Universität Hamburg, Luruper Chaussee 149, 22607 Hamburg, Germany}

\begin{abstract}
We analyze the excesses at 95~GeV in the light Higgs-boson searches in the di-photon decay channel reported by CMS and ATLAS, which combined are at the level of three standard deviations and are compatible with the excess in the $b\bar{b}$ final state observed at LEP, together with an excess in the di-photon channel at around 152~GeV reported based on a sideband analysis. We demonstrate that these excesses can be well described in a minimally extended Georgi-Machacek (meGM) model.
This is enabled by four key features of the meGM model:
(1) a natural prediction for scalar boson masses of $\lesssim200$~GeV arising from the condition to describe both the Higgs boson signal at 125~GeV and the excesses at 95~GeV, 
(2) the prediction for a doubly charged Higgs boson that can potentially enhance the di-photon decay rates, 
(3) asymmetric $WW$ and $ZZ$ couplings to neutral scalar bosons that are induced by mild custodial symmetry breaking, and 
(4) the approximate preservation of the electroweak $\rho$ parameter to be 1 at tree level.
We show in our numerical analysis that the meGM model naturally improves the fit to the LHC data around 152~GeV when describing the excesses at 95~GeV. At the same time, the model also predicts additional light CP-odd and charged scalar bosons that can be potentially probed in future experiments, which motivates dedicated searches in the upcoming LHC runs. We also present the results of sensitivity studies for the 95~and 125~GeV Higgs-boson couplings at the HL-LHC and future $e^+e^-$ colliders, which demonstrate very interesting prospects for probing the meGM model at future colliders.

\end{abstract}

\maketitle

\newpage

%================================================================
% Main Text
%================================================================
\section{Introduction}
\label{sec:intro}

In 2012, the ATLAS and CMS collaborations discovered a Higgs boson with a mass of about $125 \gev$~\cite{ATLAS:2012yve,CMS:2012qbp}. Within the current experimental and theoretical uncertainties, the properties of this new particle are consistent with the predictions of the Standard Model (SM)~\cite{CMS:2022dwd,ATLAS:2022vkf}. However, many models of physics beyond the SM (BSM) involve a SM-like light Higgs boson that is equally well compatible with the experimental results on the detected state. While the Higgs-boson sector of the SM contains only one physical Higgs particle, BSM models often give rise to an extended Higgs-boson sector containing additional scalar particles. Correspondingly, one of the prime objectives of the current and future LHC runs is the search for new BSM Higgs bosons, which is of utmost importance for exploring the underlying physics of electroweak symmetry breaking. These additional Higgs bosons can have a mass either above or below $125 \gev$. 

CMS has performed searches for scalar di-photon resonances below $125 \gev$, where the results based on the $8 \tev$ data and the full Run~2 data at $13 \tev$ showed a local excess of $2.9\,\sigma$ at $95.4 \gev$~\cite{CMS-PAS-HIG-20-002}. Subsequently, ATLAS presented the result based on their full Run~2 dataset~\cite{ATLAS-CONF-2023-035} (in the following, we refer to their analysis with higher discriminating power, the ``model-dependent'' analysis). ATLAS found an excess with a local significance of $1.7\,\sigma$ at precisely the same mass value as the one previously reported by CMS, i.e., at $95.4 \gev$. These results were combined in \citere{Biekotter:2023oen}, neglecting possible correlations. This yields
\begin{equation}
    \mu^\mathrm{exp}_{\gamma\gamma} = \muAC^{+\dmuACp}_{-\dmuACm}\,,
    \label{mugaga}
\end{equation}
corresponding to an excess with a local significance of $\sigAC\,\sigma$ at
\begin{equation}
  m_{\phi} \equiv m_{\phi}^{\rm ATLAS+CMS} = \massAC \gev\,.
\label{massAC}
\end{equation}

About two decades ago, LEP reported a local $2.3\,\sigma$ excess in the~$e^+e^-\to Z(\phi\to b\bar{b})$ searches~\cite{LEPWorkingGroupforHiggsbosonsearches:2003ing}. Taking into account the coarse mass resolution of the $b\bar b$ final state, the LEP result is consistent with a Higgs boson with a mass of $\massAC \gev$ and a signal strength of~\cite{Cao:2016uwt,Azatov:2012bz}
\begin{align}
    \mu_{b\bar{b}}^{\rm exp} = 0.117 \pm 0.057\,.
\label{mubb}
\end{align}

Furthermore, CMS also observed another excess compatible with a mass of $\sim 95 \gev$ in the search for $pp \to \phi \to \tau^+\tau^-$~\cite{CMS:2022goy}. While this excess is most pronounced at a mass of $100\gev$ with a local significance of $3.1\,\sigma$, it is also well compatible with a mass of $\sim 95 \gev$, with a local significance of $\sim 2.6\,\sigma$. At $95 \gev$, the signal strength is determined to be
\begin{align}
    \mu^{\rm exp}_{\tau\tau} = 1.2 \pm 0.5~.
\label{mutautau}
\end{align}
ATLAS has not yet published a search in the di-tau final state in the relevant mass range. The signal strength in \refeq{mutautau} is in tension with experimental bounds from recent searches performed by CMS for the production of a Higgs boson in association with a top-quark pair or in association with a $Z$~boson, with subsequent decay into a tau pair~\cite{CMS-PAS-EXO-21-018} (for the latter channel, this depends on the nature of the potential signal at $\sim 95.4 \gev$). Furthermore, the searches performed at LEP for the process $e^+e^-\to Z(\phi\to\tau^+\tau^-)$~\cite{LEPWorkingGroupforHiggsbosonsearches:2003ing} do not support the excess in the $\tau\tau$ channel. Consequently, we will not take into account the results for the $\tau\tau$ channel in our analysis.
Analyses taking into account the combined CMS/ATLAS result in the $\gamma\gamma$ channel, together with the LEP excess in the $b\bar b$ channel can be found in \citeres{Biekotter:2023oen,Belyaev:2023xnv,Aguilar-Saavedra:2023tql,Dutta:2023cig,Ellwanger:2023zjc,Cao:2023gkc,Arcadi:2023smv,Ahriche:2023wkj,Chen:2023bqr,Dev:2023kzu,Liu:2024cbr,Cao:2024axg,Kalinowski:2024uxe,Ellwanger:2024txc,Ellwanger:2024vvs,Arhrib:2024zsw,Benbrik:2024ptw,Lian:2024smg,Ghosh:2025ebc,Du:2025eop,Benbrik:2025hol} (see also \citeres{Biekotter:2017xmf,Domingo:2018uim,Biekotter:2019kde,Biekotter:2021qbc,Biekotter:2022jyr,Biekotter:2023jld} for some early analyses of the excesses in the $\gamma\gamma$ and $b \bar b$ channels).

In \citere{Chen:2023bqr}, we demonstrated that the Georgi-Machacek (GM) model~\cite{Georgi:1985nv,Chanowitz:1985ug} can very well accommodate the excesses found at about $\sim 95.4 \gev$ (some related works include Refs.~\cite{Ahriche:2023wkj,Kundu:2024sip,Ghosh:2025ebc,Du:2025eop,Mondal:2025tzi}). The GM model extends the SM with Higgs triplet fields, naturally leading to doubly charged Higgs bosons, which can potentially have an important impact on BR($\phi \to \gamma\gamma$) for BSM Higgs bosons. As we will discuss below, this feature is particularly relevant for the description of the observed excesses. One prominent feature of the GM model is that it respects the custodial symmetry and has $\rho^{\rm tree} = 1$ by construction. This is realized as a consequence of a global $SO(4)\cong SU(2)_L\times SU(2)_R\cong SU(2)_V\times SU(2)_A$ symmetry of the extended Higgs potential. After electroweak symmetry breaking, the custodial $SU(2)_V$ symmetry is preserved at the tree level, so that contributions to $\rho$ arise only at loop levels. The GM model has additional interesting features.  It can provide Majorana mass terms to neutrinos through the lepton number-violating couplings between the lepton fields and the complex Higgs triplet. Furthermore, it predicts the existence of several Higgs multiplets, with the mass eigenstates of two singlets ($H_1$ and $h$), one triplet ($H_3$), and one quintet ($H_5$) under the custodial symmetry. In particular, in \citere{Chen:2023bqr} it was demonstrated that the requirement of one light CP-even Higgs boson at $\sim 95 \gev$ together with a second CP-even Higgs boson at $\sim 125 \gev$ within the GM model gives rise to the prediction that all other Higgs bosons must be lighter than $\sim 200 \gev$, which we will explain in more detail in Section~\ref{sec:numerical}.

In a series of articles~\cite{Crivellin:2021ubm,Coloretti:2023wng,Banik:2023vxa,Bhattacharya:2023lmu,Ashanujjaman:2024pky,Crivellin:2024uhc,Banik:2024ugs,Ashanujjaman:2024lnr,Ashanujjaman:2025puu,Bhattacharya:2025rfr}, possible experimental hints for a new neutral Higgs boson with a mass around $\sim 152 \gev$ were analyzed. Here it should be noted that in the papers by ATLAS and CMS where the experimental results discussed in the above articles were presented no particular excess near 152~GeV is mentioned. Instead, the possible experimental hints were claimed based on phenomenological recast studies, using in particular side-band information.
Nevertheless, these studies motivate a further exploration of the potential hints.
The type of model favored in the above-mentioned analyses is the SM augmented by a real triplet with hypercharge zero. However, several {\it ad hoc} choices for these model interpretations have to be made, particularly the choice of a pair of one neutral and one charged Higgs boson with a mass around $\sim 152 \gev$. On the other hand, in our GM model interpretation of the $95 \gev$ excesses, such a configuration is naturally favored. The question arises whether in a GM(-type) model describing the $95 \gev$ excesses also naturally yields a description of the experimental signatures possibly hinting towards a $\sim 152 \gev$ Higgs boson (as analyzed in the above-mentioned series of articles). In the present article, we focus on the analysis performed in \citere{Ashanujjaman:2024lnr}. In this work, the decays of the possible state at $\sim 152 \gev$ to $\gamma\gamma$ together with several accompanying experimental signatures are investigated. We demonstrate that a simple, well-motivated extension of the GM model naturally leads to a significant improvement of the description of the experimental data in comparison to the SM. As a remark, we note that Ref.~\cite{Ashanujjaman:2025puu} has explored the excesses at 152~GeV in the context of the extended GM model, while this study goes beyond that by also considering the excesses at 95~GeV.

The paper is organized as follows. After a brief review of the extended GM model and its theoretical constraints in \refse{sec:eGM}, we list all relevant experimental constraints in \refse{sec:numerical}. In particular, in \refse{subsec:152} we give a description of the excesses around $\sim 152 \gev$ that we take into account. Our analysis flow is described in \refse{subsec:scan}. The main results of our analysis are presented in \refse{sec:anomaly}, and the future prospects for probing the considered scenario of an extended GM model at the HL-LHC or a future $e^+e^-$ collider are discussed in \refse{sec:prospects}. Our conclusions can be found in \refse{sec:conclusions}.  Formulas of all the Higgs boson masses are collected in Appendix~\ref{sec:mass}.

%================================================================
\section{The extended Georgi-Machacek model}\label{sec:eGM}

The Higgs sector of the extended GM model, called eGM model in the following, is composed of an isospin doublet $\phi$ with hypercharge $Y=1/2$, a complex triplet $\chi$ with $Y=1$, and a real triplet $\xi$ with $Y=0$, which are given in the $SU(2)_L$ fundamental and adjoint representations, respectively, by
\begin{equation}
    \phi = \begin{pmatrix}
        \phi^+ \\[1ex]
        \phi^0
    \end{pmatrix} ,~ \chi = \begin{pmatrix}
        \frac{\chi^+}{\sqrt{2}} & -\chi^{++} \\[1ex]
        \chi^0 & -\frac{\chi^+}{\sqrt{2}}
    \end{pmatrix} ,~ \xi = \begin{pmatrix}
        \frac{\xi^0}{\sqrt{2}} & -\xi^+ \\[1ex]
        -\xi^- & - \frac{\xi^0}{\sqrt{2}}
    \end{pmatrix} ~.
\end{equation}
The neutral components are parametrized as
\begin{equation}
    \phi^0 = \frac{1}{\sqrt{2}}(v_\phi+\phi_r+i\phi_i) ,~ \chi^0 =  v_\chi + \frac{1}{\sqrt{2}}(\chi_r+i\chi_i) ,~ \xi^0 =  v_\xi + \xi_r ~,
\end{equation}
with $v_\phi$, $v_\chi$, and $v_\xi$ denoting the vacuum expectation values (VEVs) of the corresponding fields. Without loss of generality, we can take these VEVs to be real and positive by rephasing the scalar fields~\cite{Chen:2023ins}. The Fermi constant $G_{\rm F}$ and the electroweak $\rho$-parameter at tree level satisfy
\begin{equation}\label{eq:vev}
    v^2 \equiv v_\phi^2 + 4(v_\chi^2+v_\xi^2) = \frac{1}{\sqrt{2}G_{\rm F}} ,~ \rho = \frac{v^2}{v^2 + 4(v_\chi^2-v_\xi^2)} ~.
\end{equation}
The current global fit on $\rho$ is given by the Particle Data Group as~\cite{ParticleDataGroup:2024cfk}\footnote{If instead of the world average for the $W$-boson mass, the result reported by the CDF-II Collaboration~\cite{CDF:2022hxs} were used, the value for $\rho$ would be significantly affected, see e.g.\ \citere{Chen:2022ocr}. In the present work, we choose to stick to the global fit value.}
\begin{equation}\label{eq:rho}
    \rho = 1.00031\pm0.00019 ~,
\end{equation}
which translates to
\begin{equation}\label{eq:Deltavsq}
    v_\chi^2-v_\xi^2 = (4.6969\pm2.8779)~{\rm GeV}^2 ~.
\end{equation}

The most general Higgs potential that is consistent with the electroweak symmetry is given by~\cite{Blasi:2017xmc,Keeshan:2018ypw,Chen:2023ins}
\begin{equation}\label{eq:pot_gen}
\begin{aligned}
V(\phi,\chi,\xi)
=&\,
m_\phi^2 (\phi^\dagger \phi)
+ m_\chi^2\Tr(\chi^\dagger\chi)
+ m_\xi^2\Tr(\xi^2) 
+ \mu_{\phi\xi}\phi^\dagger \xi\phi 
+ \mu_{\chi\xi}\Tr(\chi^\dagger \chi \xi)
+\lambda (\phi^\dagger \phi)^2 
\\
&
+\rho_1\left[\Tr(\chi^\dagger\chi)\right]^2+\rho_2\Tr(\chi^\dagger \chi\chi^\dagger \chi)
+\rho_3\left[\Tr(\xi^2)\right]^2
+\rho_4 \Tr(\chi^\dagger\chi)\Tr(\xi^2)
\\
&
+\rho_5\Tr(\chi^\dagger \xi)\Tr(\xi \chi)
+\sigma_1\Tr(\chi^\dagger \chi)\phi^\dagger \phi+\sigma_2 \phi^\dagger \chi\chi^\dagger \phi
+\sigma_3\Tr(\xi^2)\phi^\dagger \phi
\\
&
+ \left( \mu_{\phi\chi}\phi^\dagger \chi \tilde{\phi} + \rm{H.c.} \right)
+ \left( \sigma_4 \phi^\dagger \chi\xi \tilde{\phi} + \rm{H.c.} \right)
~,
\end{aligned}
\end{equation}
where $\mu_{\phi\chi}$ and $\sigma_4$ are generally complex parameters, and $\tilde{\phi} = i\tau^2\phi^*$ is the charge conjugation of $\phi$ with $\tau^a$ ($a=1,2,3$) being the Pauli matrices. Note that these couplings are set at the electroweak scale and that we only consider leading-order contributions throughout the analysis.
In this study, we do not consider possible imaginary parts of the couplings, \ie, we set $\Im\mu_{\phi\chi}=\Im\sigma_4=0$, and only study the CP-conserving limit of the eGM model. For a study of CP violation in this model, see \citere{Chen:2023ins}. 

The tadpole conditions, $\partial V/\partial X = 0$ for $X=\phi_r,\chi_r,\xi_r$, respectively, lead to the relations
\begin{align}\label{eq:tadpole}
m_\phi^2 &= -v_\phi^2\lambda - v_\chi[2\Re\mu_{\phi \chi} + v_\chi(\sigma_1 + \sigma_2)] + \frac{v_\xi}{\sqrt{2}}(\mu_{\phi\xi} - \sqrt{2} v_\xi\sigma_3) - \sqrt{2} v_\chi v_\xi\Re\sigma_4 ~, 
\\
m_\chi^2 &= -\frac{v_\phi^2}{4v_\chi}\Re(2\mu_{\phi\chi}  + \sqrt{2}v_\xi\sigma_4) -\frac{v_\xi}{\sqrt{2}}\mu_{\chi\xi} - \frac{v_\phi^2}{2}(\sigma_1 + \sigma_2) - 2v_\chi^2(\rho_1 + \rho_2) -v_\xi^2\rho_4 ~, 
\\
m_\xi^2 &= \frac{1}{4\sqrt{2}v_\xi}(v_\phi^2\mu_{\phi\xi} -2v_\chi^2\mu_{\chi\xi} - 2v_\phi^2v_\chi\Re\sigma_4) -\frac{v_\phi^2}{2}\sigma_3 - v_\chi^2 \rho_4 - 2v_\xi^2 \rho_3 ~.
\end{align}
Accordingly, we can derive the mass eigenstates of the scalar fields as
\begin{align}
\begin{split}
&
\chi^{\pm\pm}
=	H^{\pm\pm} ,~
{\bm H}_{\rm weak}^+
= O_{G^\pm} \tilde{\bm H}_{\rm mass}^+
= O_{G^\pm} U_{H^\pm} {\bm H}_{\rm mass}^+ ~,\\
&
{\bm H}_{\rm weak}^0
= O_{G^0} \tilde{\bm H}_{\rm mass}^0
= O_{G^0}O_{H^0}{\bm H}_{\rm mass}^0 ~, 
\end{split}
\label{eq:masseigen1}
\end{align}
where 
\begin{align}
\begin{split}
&
{\bm H}_{\rm weak}^\pm 
= 
\begin{pmatrix}
\phi^\pm \\ \chi^\pm \\ \xi^\pm
\end{pmatrix}
,~
\tilde{\bm H}_{\rm mass}^\pm 
= 
\begin{pmatrix}
\tilde{H}_1^\pm \\ \tilde{H}_2^\pm \\ G^\pm
\end{pmatrix}
,~
{\bm H}_{\rm mass}^\pm 
= 
\begin{pmatrix}
H_1^\pm \\ H_2^\pm \\ G^\pm
\end{pmatrix}
~,
\\[1ex]
&
{\bm H}_{\rm weak}^0 
= 
\begin{pmatrix}
\phi_r \\ \chi_r \\ \xi_r \\ \phi_i \\ \chi_i
\end{pmatrix}
,~
\tilde{\bm H}_{\rm mass}^0 
= 
\begin{pmatrix}
\tilde{H}_0 \\ \tilde{H}_1 \\ \tilde{H}_2 \\ \tilde{H}_3 \\ G^0
\end{pmatrix}
,~
{\bm H}_{\rm mass}^0 
= 
\begin{pmatrix}
H_0 \\ H_1 \\ H_2 \\ H_3 \\ G^0
\end{pmatrix}
~, 
\end{split} \label{eq:masseigen}
\end{align}
with 
\begin{align}
{\bm H}_{\rm weak}^-=({\bm H}_{\rm weak}^+)^* ,~
\tilde{{\bm H}}_{\rm mass}^-=(\tilde{{\bm H}}_{\rm mass}^+)^* ,~
{\bm H}_{\rm mass}^-=({\bm H}_{\rm mass}^+)^* ~.
\end{align}
Here, $G^{\pm(0)}$ denote the (would-be) Nambu-Goldstone bosons (NGBs) to be absorbed into the longitudinal components of $W^\pm(Z)$, and $H^{\pm\pm}$, $H_i^\pm$ ($i=1,2$), and $H_j$ ($j=0,\cdots,3$) the physical doubly-charged, singly-charged, and neutral Higgs bosons, respectively. We identify $H_0\equiv h$ as the 125-GeV Higgs boson discovered at the LHC and name the remaining fields in the order of their mass hierarchy, \ie, $m_{H_i}^{(\pm)}\geq m_{H_j}^{(\pm)}$ for $i>j>0$. The matrices $O_{G^\pm}$, $O_{G^0}$, and $O_{H^0}$ ($U_{H^\pm }$) are orthogonal (unitary) matrices, with the former two separating the NGB modes from the physical Higgs bosons and given simply in terms of the Higgs VEVs as 
\begin{align}
O_{G^\pm} & = \left(
\begin{array}{ccc}
  -\frac{2\sqrt{v_\xi^2+v_\chi^2}}{v} & 0 &  \frac{v_\phi}{v} \\
  \frac{v_\phi v_\chi}{v\sqrt{v_\xi^2+v_\chi^2}} & \frac{v_\xi}{\sqrt{v_\xi^2+v_\chi^2}} &  \frac{2 v_\chi}{v} \\
\frac{v_\xi v_\phi}{v\sqrt{v_\xi^2+v_\chi^2}} & -\frac{v_\chi}{\sqrt{v_\xi^2+v_\chi^2}} &  \frac{2 v_\xi}{v}  
\end{array}
\right)
,~
O_{G^0}  =  
\begin{pmatrix}
\mathbbm{1}_{3\times 3} & 0 & 0 \\ 
0& -\frac{2\sqrt{2}v_\chi}{\sqrt{v_\phi^2+8v_\chi^2}} & \frac{v_\phi}{\sqrt{v_\phi^2+8v_\chi^2}} &   \\
0& \frac{v_\phi}{\sqrt{v_\phi^2+8v_\chi^2}}           & \frac{2\sqrt{2}v_\chi}{\sqrt{v_\phi^2+8v_\chi^2}} 
\end{pmatrix} ~.
\end{align}
On the other hand, the matrices $U_{H^\pm}$ and $O_{H^0}$ are not determined purely by the VEVs but also depend on the mass matrices for the physical states. Details of the mass mixing can be found in Appendix~\ref{sec:mass}. In order to simplify the expressions, we rewrite elements of the mixing matrices defined in Eq.~(\ref{eq:masseigen}) as
\begin{align}
\begin{split}
&R_{\phi_r i} \equiv (O_{G^0}O_{H^0})_{1,i+1} ,~
R_{\chi_r i} \equiv (O_{G^0}O_{H^0})_{2,i+1} ,~
R_{\xi_r i} \equiv (O_{G^0}O_{H^0})_{3,i+1} ~, 
\\
&R_{\phi_i i} \equiv (O_{G^0}O_{H^0})_{4,i+1} ,~
R_{\chi_i i} \equiv (O_{G^0}O_{H^0})_{5,i+1} ~, 
\\
&R_{\phi^\pm j} \equiv (O_{G^\pm} U_{H^\pm})_{1j} ,~
R_{\chi^\pm i} \equiv (O_{G^\pm} U_{H^\pm})_{2j} ,~
R_{\xi^\pm j} \equiv (O_{G^\pm} U_{H^\pm})_{3j} ~, 
\end{split}
\end{align}
where $i (=0,1,2,3)$ and $j (=1,2)$ label the physical neutral and singly-charged Higgs bosons, respectively, with $H_0 \equiv h$.

The most general Yukawa interactions can be divided into two parts,
\begin{align}
{\cal L}_Y &= {\cal L}_Y^\phi + {\cal L}_Y^\chi ~, 
\end{align}
where, in the original scalar field basis,
\begin{align}
\begin{split}
{\cal L}_Y^\phi &=-y_u\bar{Q}_L\tilde{\phi}u_R - y_d\bar{Q}_L\phi d_R -y_e\bar{L}_L \phi e_R  + {\rm H.c.} ~, 
\\
{\cal L}_Y^\chi & = -y_\nu \overline{L_L^c}(i\tau_2)\chi L_L + {\rm H.c.} ~, 
\end{split}
\end{align}
with $y_{u,d,e,\nu}$ denoting the Yukawa coupling matrices for the up-type quarks, down-type quarks, and charged leptons, and the Majorana coupling matrix for the neutrinos, respectively, and $Q_L$, $L_L$, $u_R$, $d_R$, $e_R$ denoting the left-handed quark doublet, lepton doublet, right-handed up-type quark singlet, down-type quark singlet, and charged lepton singlet, respectively. The field $L_L^c=-i\gamma_2L_L^*$ is the charge-conjugated lepton doublet. The Yukawa interactions for $\phi$, ${\cal L}_Y^\phi$, take the same form as that in the SM to provide masses for the quarks and the charged leptons, while ${\cal L}_Y^\chi$ provides tiny neutrino mass via the type-II seesaw mechanism~\cite{Cheng:1980qt,Schechter:1980gr,Magg:1980ut,Mohapatra:1980yp}. Although the form of ${\cal L}_Y^\phi$ is the same as in the SM, the interaction terms between fermions and the Higgs boson are different from the SM ones due to the Higgs field mixing and are given by
\begin{align}
{\cal L}_Y^\phi 
\supset& \, 
-\sum_{f = u,d,\ell}\sum_{i = 1}^5\bar{f}\frac{M_f}{v_\phi}\left( R_{\phi_r i} -2iI_f\gamma_5 R_{\phi_i i} \right) f({\bm H}_{\rm mass}^0)_i  
\notag\\
&-\frac{\sqrt{2}}{v_\phi}\left[\bar{u}(V_{ud} M_d P_R - M_u V_{ud}P_L) d 
+ \bar{\nu}_\ell (M_\ell P_R) \ell \right] \sum_{j = 1}^3R_{\phi^\pm j} ({\bm H}_{\rm mass}^+)_j + {\rm H.c.} ~, \label{eq:yuk}
\end{align}
where $M_f (\equiv y_f v_\phi/\sqrt{2})$ is the mass of the charged fermion $f$, $I_f$ is the third component of its isospin, \ie, $I_{u}\,(I_{d,e}) = 1/2~(-1/2)$, and $V_{ud}$ is the associated element of the Cabibbo-Kobayashi-Maskawa matrix. In Eq.~\eqref{eq:yuk}, the generation indices are not explicitly shown. In the CP-conserving limit, the scalar bosons couple only through either the scalar- or pseudoscalar-type interactions.

For later convenience, here we define the coupling modifiers for the CP-even mass eigenstates $H_i$ $(i=0,1,3)$ 
\begin{equation}
\begin{aligned}\label{eq:kappa}
    \kappa_{H_iff} &= R_{\phi_r i}\frac{v}{v_\phi} ~, \\
    \kappa_{H_iWW} &= R_{\phi_r i}\frac{v_\phi}{v} + 2\sqrt{2}R_{\chi_r i}\frac{v_\chi}{v} + 4R_{\xi_r i}\frac{v_\xi}{v} ~, \\
    \kappa_{H_iZZ} &= R_{\phi_r i}\frac{v_\phi}{v} + 4\sqrt{2}R_{\chi_r i}\frac{v_\chi}{v} ~, \\
    \kappa_{H_i\gamma\gamma} &= \left(\frac{\Gamma_{H_i\to\gamma\gamma}^{\rm eGM}}{\Gamma_{H_i\to\gamma\gamma}^{\rm SM}}\right)^{1/2} ~,
\end{aligned}
\end{equation}
where $\Gamma_{H_i\to\gamma\gamma}^{\rm eGM}$ denotes the one-loop di-photon decay width of $H_i$ predicted by the eGM model and $\Gamma_{H_i\to\gamma\gamma}^{\rm SM}$ denotes the decay width of a SM-like Higgs boson with the same mass as $H_i$.

%----------------------------------------------------------------
\subsection{The custodial symmetric limit}\label{subsec:GM}

In the limit where an additional global $SU(2)_L\times SU(2)_R$ symmetry is imposed, one can express the scalar fields in terms of a bi-doublet $\Phi$ and a bi-triplet $\Delta$ as
\begin{equation}
	\Phi=\begin{pmatrix}
		\phi^{0*} & \phi^+ \\
		-\phi^- & \phi^0
	\end{pmatrix} , ~
	\Delta=\begin{pmatrix}
		\chi^{0*} & \xi^+ & \chi^{++} \\
		-\chi^- & \xi^0 & \chi^+ \\
		\chi^{--} & -\xi^- & \chi^0
	\end{pmatrix} ~.
\end{equation}
Correspondingly, the most general potential that is invariant under the $SU(2)_L\times SU(2)_R$ symmetry is given by
\begin{equation}\label{eq:VGM}
\begin{aligned}
	V(\Phi,\Delta) &= \frac{m_1^2}{2}\Tr(\Phi^\dagger\Phi)+\frac{m_2^2}{2}\Tr(\Delta^\dagger\Delta)+\lambda_1\big[\Tr(\Phi^\dagger\Phi)\big]^2+\lambda_2\big[\Tr(\Delta^\dagger\Delta)\big]^2 \\
	&\quad +\lambda_3\Tr\big[(\Delta^\dagger\Delta)^2\big]+\lambda_4\Tr(\Phi^\dagger\Phi)\Tr(\Delta^\dagger\Delta)+\lambda_5\Tr(\Phi^\dagger\frac{\tau^a}{2}\Phi\frac{\tau^b}{2})\Tr(\Delta^\dagger T^a\Delta T^b) \\
	&\quad +\mu_1\Tr \left(\Phi^\dagger\frac{\tau^a}{2}\Phi\frac{\tau^b}{2} \right)(P^\dagger\Delta P)_{ab}+\mu_2\Tr(\Delta^\dagger T^a\Delta T^b)(P^\dagger\Delta P)_{ab} ~,
\end{aligned}
\end{equation}
where $T^a$ are the $3\times 3$ matrix representations of the $SU(2)$ generators and $P$ is the similarity transformation relating the triplet and adjoint representations of the $SU(2)$ generators given by
\begin{equation}
	P=\frac{1}{\sqrt{2}}\begin{pmatrix}
		-1 & i & 0 \\
		0 & 0 & \sqrt{2} \\
		1 & i & 0
	\end{pmatrix} ~.
\end{equation}
If the triplet VEVs are aligned as 
\begin{equation}
\langle\Delta\rangle = {\rm diag}(v_\chi,v_\chi,v_\chi) ~, 
\end{equation}
the $SU(2)_L\times SU(2)_R \simeq SU(2)_V\times SU(2)_A$ symmetry is spontaneously broken down to the custodial $SU(2)_V$ symmetry. As has been pointed out in \citere{Chen:2022ocr}, the condition $\langle\chi^0\rangle=\langle\xi^0\rangle$ is necessary for the case with the $SU(2)_L\times SU(2)_R$ symmetry in order to avoid two additional charged NGB modes associated with the spontaneous breakdown of $SU(2)_V \to U(1)_{\Delta}$, the latter being an overall phase rotation symmetry of the triplet VEV. Up to this point, we have recovered the original GM model~\cite{Georgi:1985nv,Chanowitz:1985ug} from the general eGM model, which is achieved through the following identifications of the potential parameters,
\begin{align}
\begin{split}
&m_\phi^2=m_1^2,~ m_\chi^2 = m_2^2,~ m_\xi^2= \frac{m_2^2}{2},~
\mu_{\phi\xi} = -\frac{\mu_1}{\sqrt{2}},~ \mu_{\phi\chi}= \frac{\mu_1}{2},~
\mu_{\phi\chi}=6\sqrt{2}\mu_2 ~,\\
&\lambda=4\lambda_1,~ 
\rho_1=4\lambda_2+6\lambda_3,~ 
\rho_2=-4\lambda_3,~ 
\rho_3=\lambda_2+\lambda_3,~
\rho_4=4\lambda_2,~
\rho_5=4\lambda_3 ~, \\
& \sigma_1 =4\lambda_4-\lambda_5,~ 
 \sigma_2 =2\lambda_5,~ 
 \sigma_3 =2\lambda_4,~
 \sigma_4 =\sqrt{2}\lambda_5 ~. 
\end{split}\label{rel} 
\end{align}

In \citere{Chen:2023bqr}, it was shown that the GM model can accommodate the excess observed at about 95~GeV in the $\gamma\gamma$ channel by both CMS and ATLAS and by LEP in the $b\bar b$ final state (see the discussion in \refse{sec:intro}). However, while the requirement of one CP-even Higgs boson at $\sim 95 \gev$, together with a SM-like Higgs boson at $\sim 125 \gev$ naturally yields the prediction of additional Higgs bosons around $\sim 150 \gev$ (and thus a possible explanation for the excess at about 152~GeV discussed in \citeres{Crivellin:2021ubm,Bhattacharya:2023lmu,Ashanujjaman:2024pky,Crivellin:2024uhc,Banik:2024ugs,Ashanujjaman:2024lnr,Ashanujjaman:2025puu}), a problem for this interpretation occurs: since the data supporting the excess at about 152~GeV seems to suggest that a possible 152~GeV state couples much more strongly to $WW$ than to $ZZ$, the $SU(2)_V$ symmetry of the GM model obviously prevents one from incorporating such a 152~GeV resonance into its scalar spectrum. On the other hand, the eGM model is not subject to this constraint, as the $SU(2)_V$ symmetry is not imposed in the eGM model and thus the scalar couplings to $WW$ and $ZZ$ can take on more flexible values in this case. Therefore, we now turn to a minimal extension of the GM model that is a subclass of the general eGM model, in which the previous limitation is alleviated.

%----------------------------------------------------------------
\subsection{A minimal extension}\label{subsec:min}

In this section, we propose a minimal extension with respect to the original GM model discussed in Sec.~\ref{subsec:GM} that breaks the custodial symmetry and thus allows the scalar bosons to couple differently to the SM fermions and gauge bosons. The potential in this minimally extended model is defined as
\begin{align}
V_{\rm Min} &= V(\Phi,\Delta) + V_{\rm Soft} + (\sigma_4 \phi^\dagger \chi\xi \tilde{\phi} + {\rm H.c.}) ~, \label{pot_min}
\end{align}
where the first term is the $SU(2)_L\times SU(2)_R$-invariant potential given in Eq.~\eqref{eq:VGM}, and the second term, explicitly given by
\begin{align}\label{eq:soft_breaking}
V_{\rm Soft} = m_\chi^2\Tr(\chi^\dagger\chi)+m_\xi^2\Tr(\xi^2) + 
(\mu_{\phi\chi}\phi^\dagger \chi \tilde{\phi} + {\rm H.c.}) +\mu_{\phi\xi}\phi^\dagger \xi\phi + \mu_{\chi\xi}\Tr(\chi^\dagger\chi\xi) ~,
\end{align}
contains all the possible soft-breaking terms for the $SU(2)_L\times SU(2)_R$ symmetry. We note here that the last term in Eq.~\eqref{pot_min} and the third term in Eq.~\eqref{eq:soft_breaking} were originally introduced in Ref.~\cite{Chen:2023ins} to provide one single effective CP-violating degree of freedom since they are related to each other through the degenerate minimization conditions regarding $\phi_i$ and $\chi_i$. Here, we remark again that we set $\Im\mu_{\phi\chi}=\Im\sigma_4=0$ and only study the CP-conserving limit of the model in this work. As the dimension-2 and -3 terms are equivalent to those in the most general case defined in Eq.~\eqref{eq:pot_gen}, we can reparameterize the coefficients of these vertices as in Eq.~\eqref{eq:pot_gen}, \eg, $(m_2^2 + m_\chi^2)\Tr(\chi^\dagger\chi) \to m_\chi^2\Tr(\chi^\dagger\chi)$. The minimally extended model that we will study below, dubbed the ``meGM'' model in this paper, is obtained by taking the following relations of the most general case,
\begin{align}
\begin{split}
&\lambda=4\lambda_1,~ 
\rho_1=4\lambda_2+6\lambda_3,~ 
\rho_2=-4\lambda_3,~ 
\rho_3=\lambda_2+\lambda_3,~
\rho_4=4\lambda_2,~
\rho_5=4\lambda_3 ~, \\
& \sigma_1 =4\lambda_4-\lambda_5,~ 
 \sigma_2 =2\lambda_5,~ 
 \sigma_3 =2\lambda_4 ~. 
\end{split}\label{rel1}
\end{align}
The mass formulas for the Higgs bosons can be obtained by substituting the above equations into those for the general case discussed in Sec.~\ref{sec:eGM}. We remark that since the meGM model is obtained by choosing the potential parameters of the eGM model according to Eq.~\eqref{rel1} rather than eliminating certain operators based on additional symmetries, there should be no concern regarding its renormalizability.

In our global fit and numerical studies, we choose the following set as free input parameters,
\begin{equation}\label{eq:input}
	\{v_\chi,v_\xi,\lambda_2,\lambda_3,\sigma_4,\mu_{\phi\chi},\mu_{\phi\xi},\mu_{\chi\xi}\} ~,
\end{equation}
while $v_\phi$, $\lambda_1$, $\lambda_4$, and $\lambda_5$ are fixed by the requirements that the VEV $v\simeq 246$~GeV, and the Higgs boson masses are obtained as $m_h \simeq 125$~GeV, $m_{H_1} \simeq 95$~GeV, and $m_{H_3}\simeq 152$~GeV. It should be noted that we do not identify $H_2$ with any of the three above-mentioned resonances because $H_2$ is mostly composed of the CP-odd fields $\phi_i$ and $\chi_i$ according to our numerical studies, as is the case for the original GM model as discussed in \citere{Chen:2023bqr}. The masses of the additional Higgs bosons are determined by fixing the above Lagrangian parameters, where we impose the hierarchies as $m_{H_1^\pm}\leq m_{H_2^\pm}$ and $m_{H_1} \leq m_h \leq m_{H_2} \leq m_{H_3}$.

%----------------------------------------------------------------
\subsection{Theoretical constraints}\label{subsec:theoretical}

The potential parameters given in Eq.~\eqref{eq:pot_gen} can be further constrained by the following conditions: vacuum stability, boundedness from below, and perturbative unitarity. The first and third conditions for the most general eGM potential were first studied in Ref.~\cite{Chen:2023ins}, in which details of the derivations are also provided, while the second condition has been studied in Ref.~\cite{Moultaka:2020dmb}. Here we only summarize the results.

%****************************************************************
\subsubsection{Vacuum stability}\label{subsubsec:unique}

In general, it is possible that the desired electroweak vacuum $\vec{v}=(v_\phi,v_\chi,v_\xi)$ satisfying Eq.~\eqref{eq:vev} is not the global minimum of the Higgs potential and that there exist some other deeper minima. While metastable configurations where the tunneling rates into all deeper minima are so low that the lifetime of the metastable electroweak vacuum is longer than the age of the universe are phenomenologically viable, see e.g.~\citeres{Hollik:2018wrr,Ferreira:2019iqb,Biekotter:2021rak}, for simplicity we apply here the stricter criterion that the electroweak vacuum is required to be the global minimum. This can be ensured by solving for all possible VEVs that satisfy the tadpole conditions and checking whether $\vec{v}$ is indeed the global minimum. All possible VEVs can be found by solving two cubic equations of $v_\chi$ simultaneously, which are obtained from Eq.~\eqref{eq:tadpole} with $v_\phi$ and $v_\xi$ expressed in terms of the other parameters. We then check whether $\vec{v}$ is the global minimum of the scalar potential by comparing it with all the other solutions.

%****************************************************************
\subsubsection{Boundedness from below}\label{subsubsec:stability}

The Higgs potential has to be bounded from below in any direction of the field space for large field values. Such stability of the potential is ensured by the following conditions: 
\begin{align}
&	\lambda > 0,~~\rho_3 > 0,~~\rho_1 + \text{min}(\rho_2/2,\rho_2) > 0 ~, \\
&	4\lambda \rho_3 > \sigma_3^2 ,~~~4\lambda(\rho_1 + \rho_2\zeta) >  \left(\sigma_1 +  \frac{\sigma_2-\vert\sigma_2\vert\sqrt{2\zeta-1}}{2}\right)^2 ~, \\
&	4(\rho_1 + \rho_2\zeta + \rho_3 - \rho_4 -\eta\rho_5)\rho_3 > (2\rho_3 - \rho_4 - \eta\rho_5)^2 ~, \label{eq:vc3}\\
&	G(t,\zeta,\eta) > 0 ~, 
\end{align}
where 
\begin{align}
G(t,\zeta,\eta) 
\equiv & \,
4\left[ \left(\rho_1 + \rho_2\zeta + \rho_3 - \rho_4 -\eta\rho_5 \right)t^4 - (2\rho_3 - \rho_4 - \eta\rho_5)t^2 + \rho_3 \right]\lambda 
\notag\\
 & -\left[ \left(\sigma_1 + \omega\sigma_2  -\sigma_3 \right)t^2 - 2|\sigma_4|\sqrt{\frac{1-\eta}{2}} t\sqrt{1-t^2} + \sigma_3 \right]^2 ~, 
\end{align}
with the domains $t\in[0,1],~\zeta\in[1/2,1],~\eta\in[0,1]$.  We note that the condition in Eq.~\eqref{eq:vc3} is redundant if 
\begin{align}
\rho_1 + \rho_2\zeta + \rho_3 - \rho_4 -\eta\rho_5 > 0~~\text{or}~~ 2\rho_3 - \rho_4 - \eta\rho_5 > 0~\text{or}~ 0\leq \frac{ 2\rho_3 - \rho_4 - \eta\rho_5}{2(\rho_1 + \rho_2\zeta + \rho_3 - \rho_4 -\eta\rho_5) } \leq 1 ~. 
\end{align}

%****************************************************************
\subsubsection{Perturbative unitarity}\label{subsubsec:unitarity}

\begin{table}[ht!]
\centering
\begin{tabular}{c|c|c}
\toprule
$\vert Q\vert$ & $\vert Y\vert$ & Two-Body States \\
\toprule
& 0 & $\phi^+\phi^-$, $\phi^0\phi^{0*}$, $\chi^{++}\chi^{--}$, $\chi^+\chi^-$, $\chi^0\chi^{0*}$, $\xi^+\xi^-$, $\frac{\xi^0\xi^{0}}{\sqrt{2}}$ \\ \cline{2-3}
& $\frac{1}{2}$ & $\phi^{0*}\chi^0$, $\phi^-\chi^+$, $\phi^{0}\xi^0$, $\phi^-\xi^+$ \\ \cline{2-3}
0& 1 & $\frac{\phi^0\phi^0}{\sqrt{2}}$, $\chi^0\xi^0$, $\chi^+\xi^-$ \\ \cline{2-3}
& $\frac{3}{2}$ & $\phi^0\chi^0$ \\ \cline{2-3}
& 2 & $\frac{\chi^0\chi^0}{\sqrt{2}}$ \\
\colrule
 & 0 & $\phi^+\phi^{0*}$, $\chi^{++}\chi^-$, $\chi^+\chi^{0*}$, $\xi^+\xi^0$ \\ \cline{2-3}
& $\frac{1}{2}$ & $\phi^{0*}\chi^+$, $\phi^-\chi^{++}$, $\phi^0\xi^+$, $\phi^+\xi^0$, $\phi^-\chi^0$, $\phi^0\xi^-$ \\ \cline{2-3}
1& 1 & $\phi^+\phi^0$ , $\chi^{++}\xi^-$, $\chi^+\xi^{0}$, $\chi^0\xi^+$, $\chi^{0*}\xi^+$ \\ \cline{2-3}
& $\frac{3}{2}$ & $\phi^0\chi^+$, $\phi^+\chi^0$ \\ \cline{2-3}
& 2 & $\chi^+\chi^0$ \\
\colrule
 & 0 & $\frac{\xi^+\xi^+}{\sqrt{2}}$, $\chi^{++}\chi^{0*}$ \\ \cline{2-3}
& $\frac{1}{2}$ & $\phi^{0*}\chi^{++}$, $\phi^+\xi^+$ \\ \cline{2-3}
2& 1 & $\frac{\phi^+\phi^+}{\sqrt{2}}$, $\chi^{++}\xi^0$, $\chi^+\xi^+$ \\ \cline{2-3}
& $\frac{3}{2}$ & $\phi^+\chi^+$, $\phi^0\chi^{++}$ \\ \cline{2-3}
& 2 & $\chi^{++}\chi^0$, $\frac{\chi^+\chi^+}{\sqrt{2}}$ \\
\colrule
  & 1 & $\chi^{++}\xi^+$ \\ \cline{2-3}
3& $\frac{3}{2}$ & $\phi^+\chi^{++}$ \\ \cline{2-3}
 & 2 & $\chi^{++}\chi^+$ \\
\colrule
4 & 2 & $\frac{\chi^{++}\chi^{++}}{\sqrt{2}}$ \\
\botrule
\end{tabular}
\caption{\label{perturb:basis} The singlet and symmetric two-body final states formed from the doublet and triplet fields, grouped by the total electric charge ($\vert Q\vert$) and the total hypercharge ($\vert Y\vert$). A symmetry factor of $1/\sqrt{2}$ is included for the states involving identical fields.  Adapted from \citere{Chen:2023ins}.}
\end{table}

We now consider the perturbative unitarity conditions from all the high-energy $2 \to 2$ bosonic scattering processes\footnote{In this study, we only consider the perturbative unitarity conditions at tree level. For the conditions at one-loop level, we refer to Ref.~\cite{Chowdhury:2024mfu}. }.  The longitudinal modes of the weak vector bosons are taken into account as the NGB modes by using the equivalence theorem~\cite{Cornwall:1974km}. In \refta{perturb:basis}, we list all $2 \to 2$ scattering states considered, classified according to the total electric charge $Q$ and the total hypercharge $Y$.  We note that scatterings between states with different hypercharges (not only the electric charge) do not occur because the hypercharge should be conserved in the high-energy limit. We impose the following criterion for each eigenvalue $x_i$ of the $s$-wave amplitude matrix,
 \begin{equation}
	\vert\Re x_i\vert < 8\pi ~.
 \end{equation}
We find the nineteen independent eigenvalues as follows~\cite{Chen:2023ins}:
\begin{align}
\begin{split}
x_1&=2 (\rho_1+\rho_2) ~, \\
x_2&= 2\rho_1 - \rho_2 ~, \\
x_3&= 2 \rho_4+\rho_5 ~,\\
x_4&= 2 (\rho_4 + 2 \rho_5) ~, \\
x_5&= \sigma_1+\sigma_2 ~,\\
x_6&= \sigma_1 - \frac{\sigma_2}{2} ~, \\
x_7^\pm&=\rho_1+4 \rho_3 \pm \sqrt{(\rho_1-4 \rho_3)^2+2 \rho_5^2} ~,\\
x_8^\pm&=\lambda +\rho_1+2 \rho_2 \pm\sqrt{(\lambda - \rho_1 - 2 \rho_2)^2+\sigma_2^2} ~,\\
x_9^\pm&= \frac{\sigma_1}{2} + \sigma_3 \pm\frac{1}{2}\sqrt{(\sigma_1-2 \sigma_3)^2+4 |\sigma_{4}|^2} ~,\\
x_{10}^\pm&= \lambda +\rho_4-\frac{\rho_5}{2} \pm \frac{1}{2}\sqrt{(2 \lambda -2 \rho_4+\rho_5)^2+8 |\sigma_{4}|^2} ~,\\
x_{11}^\pm&=\frac{\sigma_1}{2}+\frac{3}{4} \sigma_2+\sigma_3 \pm\frac{1}{4}\sqrt{(2 \sigma_1+3 \sigma_2-4 \sigma_3)^2+64 |\sigma_{4}|^2} ~,
\end{split}
\end{align}
and $x_{12}^{i}$ $(i=1,2,3)$ are obtained as the eigenvalues of the following matrix
\begin{align}
\left(
\begin{array}{ccc}
 20 \rho_3 & 2 \sqrt{3} \sigma_3 & \sqrt{2} (3\rho_4 +\rho_5) \\
 2 \sqrt{3}\sigma_3 & 6 \lambda  & \sqrt{\frac{3}{2}} (2\sigma_1 + \sigma_2 ) \\
 \sqrt{2} (3\rho_4 + \rho_5 ) & \sqrt{\frac{3}{2}} (2\sigma_1 + \sigma_2 ) & 8\rho_1 +6\rho_2 \\
\end{array}
\right) ~. 
\end{align}
We note that the generally complex parameter $\sigma_4$ appears in the form of its absolute value in the above eigenvalues.  This is because the scattering amplitudes are evaluated in the high-energy limit, where only the quartic couplings in the potential are relevant, and the CPV phase can be removed by rephasing the scalar fields.

%================================================================
\section{Numerical analysis setup}\label{sec:numerical}

%----------------------------------------------------------------
\subsection{Experimental results for the 95-GeV excess}\label{subsec:95}

As discussed in \refse{sec:intro} (see also \citere{Chen:2023bqr}), the interpretation of the excess reported by CMS in the $\tau\tau$ channel at 95~GeV is complicated by the fact that this excess is in some tension with other searches involving the di-tau final state. Therefore, we only take into account the $b\bar{b}$ and $\gamma\gamma$ excesses in our present analysis and quantify their compatibility with our model using the quantity
\begin{equation}
    \chi^2_{95} 
    = 
    \sum_{X=\gamma\gamma,b\bar{b}}
    \left(\frac{\mu_{X,95}-\mu^{\rm exp}_{X,95}}{\Delta\mu^{\rm exp}_{X,95}}\right)^2 ~.
\end{equation}
Here, the experimental central values $\mu^{\rm exp}_{X,95}$ and uncertainties $\Delta\mu^{\rm exp}_{X,95}$ are listed in Sec.~\ref{sec:intro}, while $\mu_{X,95}$ denotes the value of the signal strength in the channel $X$ as predicted by the \megm model that we consider here. Using the framework of coupling modifiers, we can define 
\begin{align}
    \mu_{\gamma\gamma,95} &= \kappa_{H_1tt}^2\times\frac{{\rm BR}(H_1\to \gamma\gamma)}{{\rm BR}(H_1\to \gamma\gamma)_{\rm SM}} ~, \\
    \mu_{b\bar{b},95} &= \kappa_{H_1ZZ}^2\times\frac{{\rm BR}(H_1\to b\bar{b})}{{\rm BR}(H_1\to b\bar{b})_{\rm SM}} ~,
\end{align}
where we assume 100\% gluon-fusion production for $H_1$ at the LHC and denote the branching ratio for a SM-like Higgs boson at 95~GeV to the final state $X$ as ${\rm BR}(H_1\to X)_{\rm SM}$. Because $H_1$ is CP-even, we can identify $\kappa_{H_1tt}^2=[R_{\phi_r1}(v/v_\phi)]^2$ [see Eq.~\eqref{eq:kappa}].

%----------------------------------------------------------------
\subsection{Possible hints for an excess at 152~GeV}\label{subsec:152}

After reviewing the experimental situation, \citere{Ashanujjaman:2024lnr} came to the conclusion that ``statistically significant deviations from the SM predictions in final states with multiple leptons, missing energy and possibly ($b$-)jets~\cite{vonBuddenbrock:2016rmr,vonBuddenbrock:2017gvy,Buddenbrock:2019tua,vonBuddenbrock:2020ter,Hernandez:2019geu,Coloretti:2023wng,Banik:2023vxa,Coloretti:2023yyq}'' are compatible with the production of a $\approx270$~GeV Higgs boson decaying into a pair of lighter Higgs bosons, which then dominantly decay to $WW$ and $b\bar{b}$, respectively. The former of these had been studied in Ref.~\cite{vonBuddenbrock:2017gvy} and found to be consistent with a mass of $150\pm5$~GeV. Later on, Refs.~\cite{Crivellin:2021ubm,Bhattacharya:2023lmu} analyzed the sidebands of the SM Higgs boson searches conducted in Refs.~\cite{CMS:2021kom,ATLAS:2022tnm,ATLAS:2020ior,CMS:2020cga,ATLAS:2021jbf,CMS:2018nlv,CMS:2018myz,ATLAS:2020fcp} and suggested the presence of a new narrow resonance in the di-photon and $Z\gamma$ spectra with a mass of around $152$~GeV. Finally, Refs.~\cite{Ashanujjaman:2024pky,Crivellin:2024uhc,Banik:2024ftv,Ashanujjaman:2024lnr} analyzed the associated productions of the SM Higgs boson in various di-photon channels studied in Refs.~\cite{ATLAS:2023omk,ATLAS:2024lhu} and performed a fit to the data with a $\approx152$~GeV real Higgs triplet. 
As mentioned above, the possible experimental hints for a new particle at about 152~GeV are based on phenomenological recast studies rather than on the original analyses of ATLAS and CMS.

Given that the eGM model also accommodates a Higgs boson that couples asymmetrically to $WW$ and $ZZ$, in the following we analyze the question whether the meGM model introduced above, which naturally predicts a spectrum of light scalar bosons if it contains a state that is compatible with the 95~GeV excesses (see Sec.~\ref{sec:anomaly}), could also provide a possible explanation for the 152~GeV excesses.

We incorporate the experimental results reported on the 152~GeV excesses in the following way. Due to the absence of an associated excess in the $ZZ$ final states, we first require ${\rm BR}(H_3\to ZZ) < 0.1$ and ${\rm BR}(H_3\to WW) > 0.1$. Then, we follow \citere{Ashanujjaman:2024lnr} to perform a proof-of-concept sideband analysis on the results reported in  \citeres{ATLAS:2023omk,ATLAS:2024lhu} by the ATLAS Collaboration. These two studies analyze the $\gamma\gamma+X$ and $\gamma\gamma+(\ell,\tau)$ channels, respectively, targeting the production of a SM-like Higgs$+ X$. However, a possible resonance at 152~GeV in the sideband distributions of the $m_{\gamma\gamma}$ spectra could manifest itself in several signal regions.  Although this is yet to be confirmed by a dedicated experimental search, the goal of our analysis is to demonstrate that if the hints for a 152-GeV resonance are further substantiated, the \megm model considered here can naturally accommodate both the 95-GeV and the 152-GeV resonances.

\begin{figure}[ht!]
\centering
\begin{tikzpicture}
	\begin{feynman}
		\vertex (i1) {$\overline{q}^\prime$};
            \vertex at ($(i1) + (0.0cm, -2.6cm)$) (i2) {$q$};
            \vertex at ($(i1) + (1.0cm, -1.3cm)$) (w1);
            \vertex at ($(w1) + (2.0cm, 0.0cm)$) (w2);
            \vertex at ($(w2) + (1.0cm, 1.3cm)$) (H3);
            \vertex at ($(w2) + (1.0cm, -1.3cm)$) (H2pm);
            \vertex at ($(H3) + (1.0cm, 1.0cm)$) (a1) {$\gamma$};
            \vertex at ($(H3) + (1.0cm, -1.0cm)$) (a2) {$\gamma$};
            \vertex at ($(H2pm) + (1.0cm, 1.0cm)$) (X) {$X$};
            \vertex at ($(H2pm) + (1.0cm, -1.0cm)$) (Y) {$Y$};

		\diagram* {
			(i2) --[fermion] (w1) --[fermion] (i1);
                (w1) --[boson, edge label=$W^{\pm*}$] (w2);
                (w2) --[scalar, edge label=$H_3$] (H3);
                (w2) --[scalar, edge label'=$H_{1,2}^{\pm}$] (H2pm);
                (H3) --[boson] (a1);
                (H3) --[boson] (a2);
                (H2pm) -- (X);
                (H2pm) -- (Y);
		};
	\end{feynman}
\end{tikzpicture}
\caption{\label{fig:DY} A representative Feynman diagram of the charged Drell-Yan production of the scalar boson pair $H_3$ and $H_{1,2}^\pm$, which then decay to $\gamma\gamma$ and $XY=WZ,tb,cs,\tau\nu_\tau$, respectively. For $H_2^\pm$ furthermore the decay modes into $H_1^\pm Z$, $H_1W^\pm$, and $hW^\pm$ can occur.}
\end{figure}
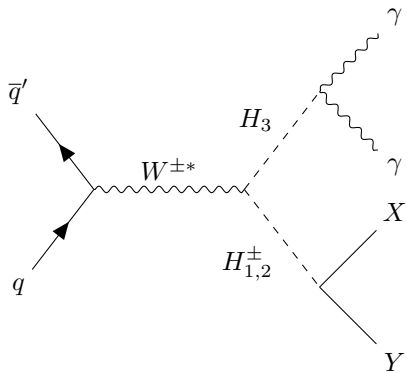

The processes we study are $qq' \to W^\pm \to H_3 H_i^\pm$, $qq' \to W^\pm \to H_3 W^\pm$, and $qq \to q'q' W^\pm W^\mp \to q'q' H_3$, where we identify the $H_3$ boson with a neutral Higgs boson with a mass of $\sim 152 \gev$ and $H_i^\pm$ with the two singly-charged  eigenstates $H_{1,2}^\pm$. For simplicity, we only consider the scenario with $m_{H_1^\pm} \approx 100 \gev$ and $m_{H_2^\pm} \approx 156 \gev$, which are the masses with the highest statistical weights in our general numerical simulations (see Sec.~\ref{sec:anomaly} and Fig.~\ref{fig:masses} below). For the decay of the neutral $H_3$ boson to $\gamma\gamma$, we find BR$(H_3 \to \gamma\gamma) \sim\mathcal{O}(10^{-3})$ (see Fig.~\ref{fig:BRH3} below). On the other hand, we consider four possible decay modes of $H_{1,2}^\pm$ into SM particles: $W^\pm Z$, $tb$, $cs$, and $\tau^\pm \nu_\tau$. In addition to those four modes, we find in our simulation that $H_2^\pm$ can also decay to $H_1^\pm Z$, $H_1 W^\pm$, and $h W^\pm$. A representative Feynman diagram of the charged Drell-Yan process is shown in \reffi{fig:DY}.

%%%%%%%%%%%%%%%%%%%%%%%%%%% F I G U R E %%%%%%%%%%%%%%%%%%%%%%%%%%%%%%%%%%%%%%%%%%%%%%%%%%%%%%%%%%%%%%%%%%
%%%%%%%%%%%%%%%%%%%%%%%%%%% F I G U R E %%%%%%%%%%%%%%%%%%%%%%%%%%%%%%%%%%%%%%%%%%%%%%%%%%%%%%%%%%%%%%%%%%

We perform the generator-level simulation with \texttt{MadGraph5\_aMC@NLO  v3.5.7}~\cite{Alwall:2014hca,Frederix:2018nkq} using the \texttt{NNPDF40\_nlo\_as\_01180} parton distribution function set, followed by the parton shower/hadronization simulation with \texttt{Pythia 8.311}~\cite{Sjostrand:2014zea} and the detector simulation with \texttt{Delphes 3.5.0}~\cite{deFavereau:2013fsa}. Jet reconstruction is performed using the anti-$k_T$ algorithm~\cite{Cacciari:2008gp} with \texttt{FastJet 3.4.2}~\cite{Cacciari:2011ma}. We follow the object and event preselection criteria of \citere{ATLAS:2023omk}, which we summarize in \refta{tab:object}. The signal regions (SR's) are chosen to capture the possible decays of the particle(s) produced together with $H_3$ ($H_{1,2}^\pm$, $W^\pm$, $qq'$). Concerning the ``relevant'' SR's, we follow \citere{Ashanujjaman:2024lnr} and study the eight regions listed in \refta{tab:SR}, but for simplicity ignore the correlations. On the other hand, we note that in 12 other potential SR's the meGM model does not yield the lowest-order contributions. In our analysis we incorporate all 20 SR's. Specifically, we obtain a $\chi^2$ value for the comparison between the predictions and the data that takes into account all SR's considered by the experimental studies in which data is available between $m_{\gamma\gamma} \in [150,155]$~GeV, including explicitly those that are not expected to have an excess at 152~GeV for the case of the meGM model.\footnote{On the contrary, \citere{Ashanujjaman:2024lnr} only accounts for the SR's in which the model considered there makes explicit contributions.} Overall, the considered 20 SR's at $151 \gev$ and $154 \gev$ result in a total of 37 bins (some signal regions have only one bin with data).

%%%%%%%%%%%%%%%%%%%%%%%%%%%%%%%%%% T A B L E %%%%%%%%%%%%%%%%%%%%%%%%%%%%%%%%%%%%%%%%%%%%%%%%%%%%%%%%%%%
\begin{table}[ht!]
    \centering
    \begin{tabular}{
    >{\centering\arraybackslash}p{3.5cm}
    |>{\centering\arraybackslash}p{6.5cm}
    }
    \toprule
    Object & Requirements \\
    \toprule
    Photon & $p_T>22~{\rm GeV},~\vert\eta\vert<2.5$ \\
    Electron & $p_T>10~{\rm GeV},~\vert\eta\vert<2.5$ \\
    Muon & $p_T>10~{\rm GeV},~\vert\eta\vert<2.7$ \\
    Jet & $p_T>25~{\rm GeV},~\vert\eta\vert<4.4$ \\
    $b$-jet & $p_T>25~{\rm GeV},~\vert\eta\vert<2.5$ \\
    $\tau$-lepton & $p_T>20~{\rm GeV},~\vert\eta\vert<2.5$ \\
    \toprule
    \multicolumn{2}{c}{Event Preselection} \\
    \toprule
    \multicolumn{2}{c}{$p_T^{\gamma_1}>35~{\rm GeV},~p_T^{\gamma_2}>25~{\rm GeV},~p_T^{\gamma_1}/m_{\gamma\gamma}>0.35,~~p_T^{\gamma_2}/m_{\gamma\gamma}>0.25$} \\
    \botrule
    \end{tabular}
    \vspace{1em}
    \caption{The object and event preselection criteria used in \protect\citere{ATLAS:2023omk}.}
    \label{tab:object}
    \vspace{2em}
\end{table}
%%%%%%%%%%%%%%%%%%%%%%%%%%%%%%%%%% T A B L E %%%%%%%%%%%%%%%%%%%%%%%%%%%%%%%%%%%%%%%%%%%%%%%%%%%%%%%%%%%

%%%%%%%%%%%%%%%%%%%%%%%%%%%%%%%%%% T A B L E %%%%%%%%%%%%%%%%%%%%%%%%%%%%%%%%%%%%%%%%%%%%%%%%%%%%%%%%%%%
\begin{table}[ht!]
    \centering
    \begin{tabular}{
    >{\centering\arraybackslash}p{3.5cm}
    |>{\centering\arraybackslash}p{10.0cm}
    }
    \toprule
    Signal Regions & Selection Criteria \\
    \toprule
    $4j$ & $n_{\rm jet}\geq 4,~\vert\eta_{\rm jet}\vert<2.5$ \\
    \colrule
    $\ell b$ & $n_{\ell=e,\mu}\geq1,~n_{b-{\rm jet}}\geq 1$ \\
    \colrule
    $t_{\rm lep}$ & $n_{\ell=e,\mu}=1,~n_{\rm jet}=n_{b-{\rm jet}}= 1$ \\
    \colrule
    $2\ell$ & $ee,\mu\mu$, or $e\mu$ \\
    \colrule
    $1\ell$ & $n_{\ell=e,\mu}=1,~n_{\tau_{\rm had}}=0,~n_{b-{\rm jet}}=0,~E_T^{\rm miss}>35~{\rm GeV}$ if $\ell=e$ \\
    \colrule
    $1\tau_{\rm had}$ & $n_{\ell=e,\mu}=0,~n_{\tau_{\rm had}}=1,~n_{b-{\rm jet}}=0,~E_T^{\rm miss}>35~{\rm GeV}$  \\
    \colrule
    $E_T^{\rm miss} > 100~{\rm GeV}$ & $E_T^{\rm miss} > 100~{\rm GeV}$ \\
    \colrule
    $E_T^{\rm miss} > 200~{\rm GeV}$ & $E_T^{\rm miss} > 200~{\rm GeV}$ \\
    \botrule
    \end{tabular}
    \vspace{1em}
    \caption{The relevant signal regions for the $qq' \to W^\pm \to H_3 H_i^\pm$, $qq' \to W^\pm \to H_3 W^\pm$, and $qq \to q'q' W^\pm W^\mp \to q'q' H_3$ processes, which are listed in \citere{Ashanujjaman:2024lnr} and were originally defined in \citeres{ATLAS:2023omk,ATLAS:2024lhu}. It should be noted that $n_{\rm jet}$ counts both light and $b$-jets.
    }
    \label{tab:SR}
    \vspace{2em}
\end{table}
%%%%%%%%%%%%%%%%%%%%%%%%%%%%%%%%%% T A B L E %%%%%%%%%%%%%%%%%%%%%%%%%%%%%%%%%%%%%%%%%%%%%%%%%%%%%%%%%%%

To quantify the compatibility of the meGM model with the di-photon excess at $\sim 152 \gev$, we define
\begin{equation}
    \chi^2_{152} 
    = 
    \sum_{{\rm SR},m_{\gamma\gamma}}
    \left(
    \frac{
    N^{{\rm exp,SR},m_{\gamma\gamma}}_{\gamma\gamma,152} -
    N_{\gamma\gamma,152}^{{\rm SR},m_{\gamma\gamma}}
    }{
    \Delta N^{{\rm exp,SR},m_{\gamma\gamma}}_{\gamma\gamma,152}}
    \right)^2~,
    \label{chi152}
\end{equation}
where $N^{{\rm exp,SR},m_{\gamma\gamma}}_{\gamma\gamma,152}$ and $\Delta N^{{\rm exp,SR},m_{\gamma\gamma}}_{\gamma\gamma,152}$ denote, respectively, the experimental central value and uncertainty of the event number in the signal region SR at the di-photon invariant mass $m_{\gamma\gamma} = 151, 154 \gev$. Furthermore,
\begin{equation}
\begin{gathered}
    N_{\gamma\gamma,152}^{{\rm SR},m_{\gamma\gamma}} = N_{\gamma\gamma,{\rm SM}}^{{\rm SR},m_{\gamma\gamma}} + k\sum_{XY} N_{\gamma\gamma,152}^{{\rm SR},m_{\gamma\gamma},XY} ~, \\
    N_{\gamma\gamma,152}^{{\rm SR},m_{\gamma\gamma},XY} = \sigma_{\rm fid}^{{\rm SR},m_{\gamma\gamma}}\times {\cal L} ~,
\end{gathered}
\end{equation}
where $N_{\gamma\gamma,{\rm SM}}^{{\rm SR},m_{\gamma\gamma}}$ is the SM background event number taken from the ATLAS simulation, $k=1.15$ is the $k$-factor used to account for the next-to-leading-order and next-to-next-to-leading-logrithmic QCD corrections~\cite{Ashanujjaman:2024lnr}, $\sigma_{\rm fid}^{{\rm SR},m_{\gamma\gamma}}$ is the sum of the fiducial cross sections of the $qq' \to W^\pm \to H_3 H_i^\pm$, $qq' \to W^\pm \to H_3 W^\pm$, and $qq \to q'q' W^\pm W^\mp \to q'q' H_3$ processes contributing to the signal region SR at $m_{\gamma\gamma}$, and we have taken the integrated luminosity to be ${\cal L} = 139~{\rm fb}^{-1}$~\cite{ATLAS:2023omk,ATLAS:2024lhu}. Although we have listed all possible channels above, we find from our numerical simulations that most of the contributions come from the following three channels: $H_2^\pm\to WZ$, $H_2^\pm\to H_1^\pm(cs)Z$, and $H_2^\pm\to H_1^\pm(\tau\nu_\tau)Z$, while the other channels only contribute marginally.

%----------------------------------------------------------------

\subsection{Parameter scan and point selection}
\label{subsec:scan}

We follow the same two-stage analysis framework as in \citere{Chen:2023bqr} and choose the parameters listed in \refeq{eq:input} as the input parameters. The first stage is to use the Bayesian-based Markov-Chain Monte Carlo simulation package \texttt{HEPfit}~\cite{DeBlas:2019ehy} to generate a collection of samples that are ``shaped'' by the applied theoretical and experimental constraints, for the latter of which we only consider the 125-GeV Higgs rate measurements from LHC Run-1 for reasons discussed in \citere{Chen:2023bqr}.\footnote{
The reason for not including the LHC Run-2 data is that some Run-2 measurements are reported in two different statistical frameworks: the signal strength measurements and the Simplified Template Cross Section measurements. While \texttt{HEPfit} adopts the former, \texttt{HiggsTools} adopts the latter, and thus, for technical convenience, we only use the Run-1 data in HEPfit, as discussed in \citere{Chen:2023bqr}.} 
The second stage is to further constrain the allowed parameter space by applying the package \texttt{HiggsTools}~\cite{Bahl:2022igd} in order to ensure that the allowed parameter regions are in accordance with the measured properties of the detected Higgs boson at 125~GeV (sub-package \texttt{HiggsSignals}~\cite{Bechtle:2013xfa,Bechtle:2014ewa,Bechtle:2020uwn,Bahl:2022igd}, dataset \texttt{v1.1}) and with the limits from searches for additional Higgs bosons at the LHC and at LEP (sub-package \texttt{HiggsBounds}~\cite{Bechtle:2008jh,Bechtle:2011sb,Bechtle:2013wla,Bechtle:2020pkv,Bahl:2022igd}, dataset \texttt{v1.2}). The employed versions of \texttt{HiggsSignals} and \texttt{HiggsBounds} include essentially all relevant datasets from the LHC Run~2 that have been made public up to now.

After applying the bounds, we then calculate the following quantities for the \megm model and the SM:
\begin{description}
    \item[$\bullet~\chi^2_{125}({\rm meGM})$] The $\chi^2$ value associated with the rate measurements of the 125-GeV Higgs boson for the individual data points, evaluated by \texttt{HiggsSignals}.
    \item[$\bullet~\chi^2_{95}({\rm meGM})$] The $\chi^2$ value associated with the 95-GeV excesses in the $\gamma\gamma,b\bar{b}$ channels for the individual data points, see \refse{sec:intro}.
    \item[$\bullet~\chi^2_{152}({\rm meGM})$] The $\chi^2$ value associated with the 152-GeV excesses in the $\gamma\gamma$ channel for the individual data points, evaluated with \refeq{chi152}.
    \item[$\bullet~\chi^2_{125}({\rm SM})$] The $\chi^2$ value associated with the rate measurements of the 125-GeV Higgs boson for the case of the SM prediction, evaluated by \texttt{HiggsSignals}.  We find $\chi^2_{125}({\rm SM}) \approx 152.5$.
    \item[$\bullet~\chi^2_{\rm 95}({\rm SM})$] The $\chi^2$ value associated with the 95-GeV excesses in the $\gamma\gamma,b\bar{b}$ channels for the case of the SM, see \refse{sec:intro}. We find $\chi^2_{95}({\rm SM}) \approx 13.2$.
    \item[$\bullet~\chi^2_{\rm 152}({\rm SM})$] The $\chi^2$ value associated with the 152-GeV excesses in the $\gamma\gamma$ channel for the case of the SM, evaluated with \refeq{chi152}. We find $\chi^2_{152}({\rm SM}) \approx 43.1$ for the total of the contributing 37 bins, indicating the level of tension between the experimental data and the SM prediction.
    \item[$\bullet~\Delta\chi^2_{M}$] The difference in the $\chi^2$ values associated with the rate measurements of the $M$-GeV Higgs boson between the meGM model and the SM, with $M=95,125,152$, which is given by $\Delta\chi^2_{M} \equiv \chi^2_{M}({\rm meGM})-\chi^2_{M}({\rm SM})$.
    \item[$\bullet~\Delta\chi^2_{95+125}$] $\Delta\chi^2_{95+125} \equiv \Delta\chi^2_{95} + \Delta\chi^2_{125}$.
    \item[$\bullet~\Delta\chi^2$] $\Delta\chi^2 \equiv \Delta\chi^2_{\rm all} \equiv \Delta\chi^2_{95} + \Delta\chi^2_{125} + \Delta\chi^2_{152}$.
\end{description}
After the numerical $\chi^2$ evaluation, we first select the subset of samples that satisfy $\Delta\chi^2_{95+125} < 0$ and $\Delta\chi^2_{125}<6.18$. For these points, the combined $\chi^2$ value arising from the excesses at 95~GeV and the measurements of the Higgs boson at 125~GeV is lower than that of the SM, while the second condition ensures the compatibility with the 125-GeV Higgs measurements by requiring that the predictions for those parameter points for the 125-GeV Higgs measurements are within the 95.4\% (or $2\,\sigma$) CL relative to the SM case for a two-dimensional parameter distribution. Next, we evaluate the subset with $m_{H_1^\pm}= 100\pm1 \gev$ and $m_{H_2^\pm}= 156\pm1 \gev$, selecting the data points with $\Delta\chi^2_{\rm all}\equiv\Delta\chi^2 < 0$. Finally, we identify the best-fit point from the latter subset.

Before closing this section, we note that a recast constraint on the singly charged Higgs boson $H^\pm$ was presented in 
\citere{Ashanujjaman:2024lnr}, which claimed to exclude charged Higgs-boson masses of $100 \gev \lesssim m_{H^{\pm}} \lesssim 110 \gev$. However, we do not consider this constraint for two reasons: First, our analysis tool for exclusion limits for BSM Higgs bosons, \texttt{HiggsBounds}, is designed to only use the bounds reported by experimental collaborations for consistency. Second, more fundamentally, in Fig.~11 of \citere{Ashanujjaman:2024lnr}, which is supposed to provide the exclusion limits on $m_{H^{\pm}}$, the  observed limit is above the model prediction for the entire mass range between 100~GeV and 200~GeV, while only the expected limit  does intersect with the model prediction at $\sim 110 \gev$. Consequently, their figure does not support the exclusion of $m_{H^{\pm}} \lesssim 110 \gev$ as concluded in \citere{Ashanujjaman:2024lnr}.

%================================================================
\section{Results}\label{sec:anomaly}

\begin{figure}[ht!]
    \centering
    \includegraphics[width=0.42\textwidth]{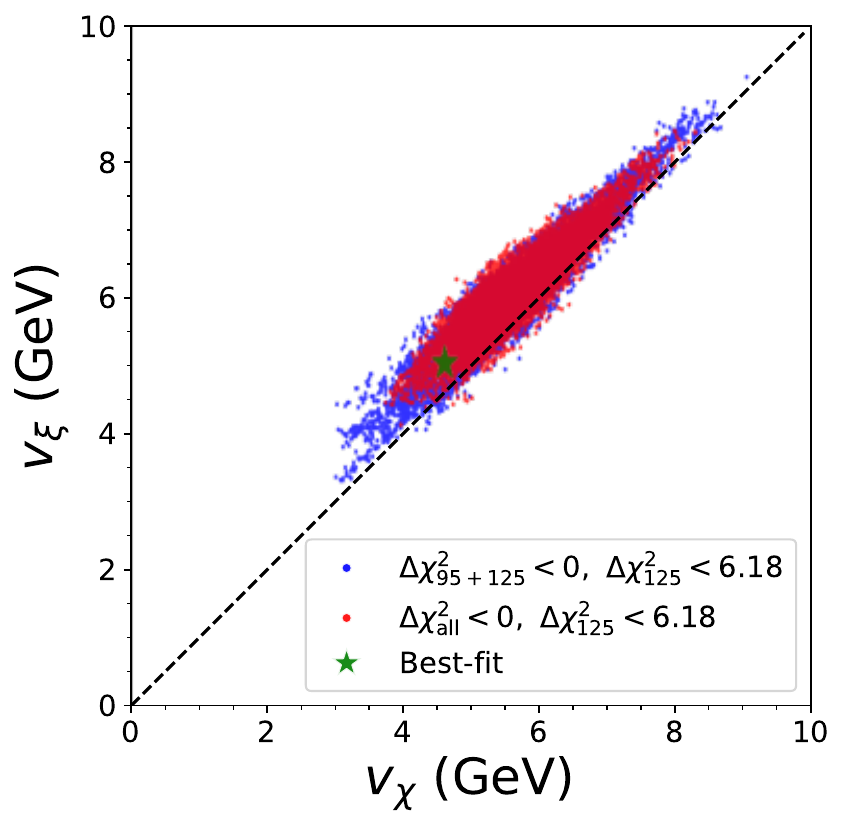}
    \includegraphics[width=0.56\textwidth]{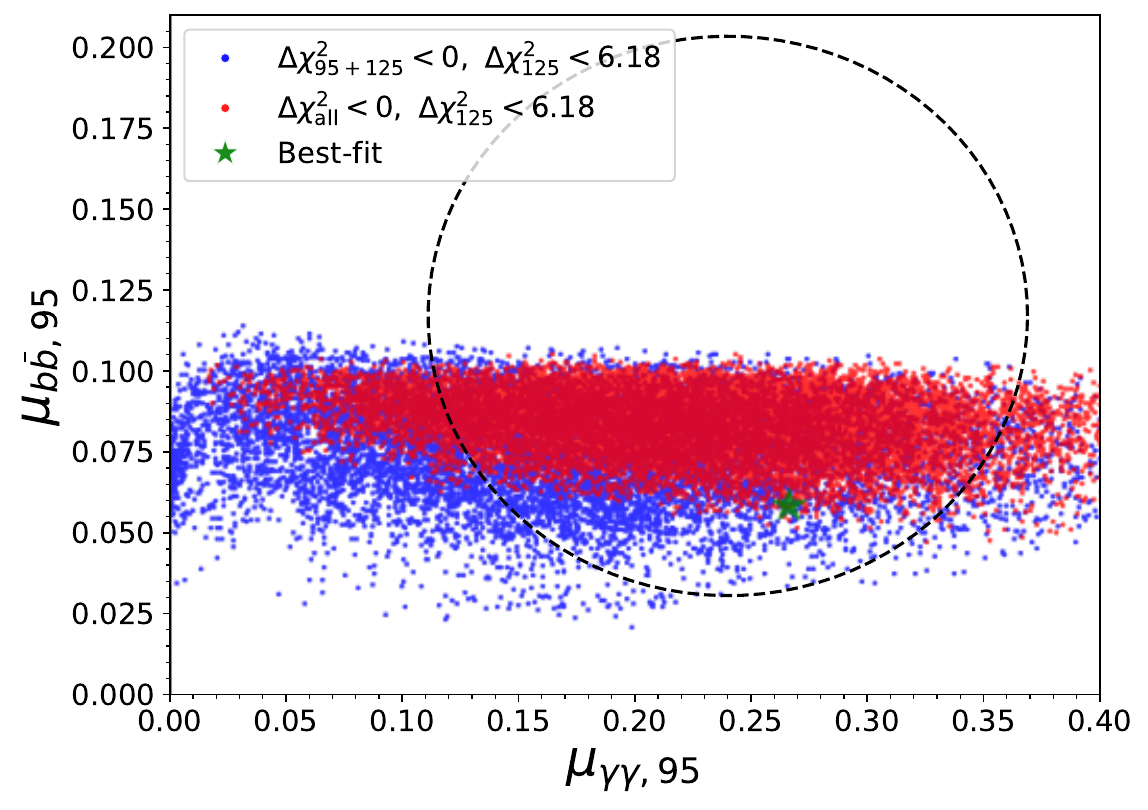}
    \caption{{\it Left:} Sample distributions in the $v_\chi$--$v_\xi$ plane, where the dashed line indicates the custodial-symmetric limit, \ie, $v_\chi=v_\xi$. {\it Right:} Sample distributions in the $\mu_{\gamma\gamma,95}$--$\mu_{b\bar{b},95}$ plane. The elliptic contour denotes the 1\,$\sigma$ bounds of the corresponding 95-GeV excess measurements. For the color coding, see the main text.}
    \label{fig:vev-mu_H1}
\end{figure}

We first present in \reffi{fig:vev-mu_H1} the sample distributions in the $v_\chi$--$v_\xi$ (left) and $\mu_{\gamma\gamma,95}$--$\mu_{b\bar{b},95}$ (right) planes. The blue points fulfill $\Delta\chi^2_{95,125} < 0$, $\Delta\chi^2_{125} < 6.18$ (the first subsample defined above). The red points fulfill $\Delta\chi^2_{95,125,152} < 0$, $\Delta\chi^2_{125} < 6.18$ as well as $m_{H_1^\pm}\approx 100 \gev$ and $m_{H_2^\pm}\approx156$~GeV, (i.e., a subset of the first subsample, corresponding to the second subsample defined above). The green star denotes the best-fit point of the latter subsample. In the left plot, the dashed line indicates the custodial-symmetric limit ($v_\chi=v_\xi$). In the right plot, we further show the (black dashed) 1\,$\sigma$ ellipse of the respective 95-GeV excesses. In the left plot, one can see that for most parameter points in the two samples $v_\xi\gtrsim v_\chi$ is preferred, as expected from the requirement of ${\rm BR}(H_3\to ZZ)\ll {\rm BR}(H_3\to WW)$ [see \refeqs{eq:rho} and (\ref{eq:kappa})], while the distribution of points is still close to the custodial-symmetric limit, as required by the constraint on the electroweak $\rho$-parameter [see \refeq{eq:Deltavsq}]. At the same time, most of the points are well within the 1\,$\sigma$ contour of the two 95-GeV excesses, confirming that the \megm model is capable of describing the 95-GeV resonance, as expected from a generalization of the GM model~\cite{Chen:2023bqr}.

%%%%%%%%%%%%%%%%%%%%%%%%%%% F I G U R E %%%%%%%%%%%%%%%%%%%%%%%%%%%%%%%%%%%%%%%%%%%%%%%%%%%%%%%%%%%%%%%%%%
%%%%%%%%%%%%%%%%%%%%%%%%%%% F I G U R E %%%%%%%%%%%%%%%%%%%%%%%%%%%%%%%%%%%%%%%%%%%%%%%%%%%%%%%%%%%%%%%%%%

\begin{figure}[ht!]
    \centering
    \includegraphics[width=0.48\textwidth]{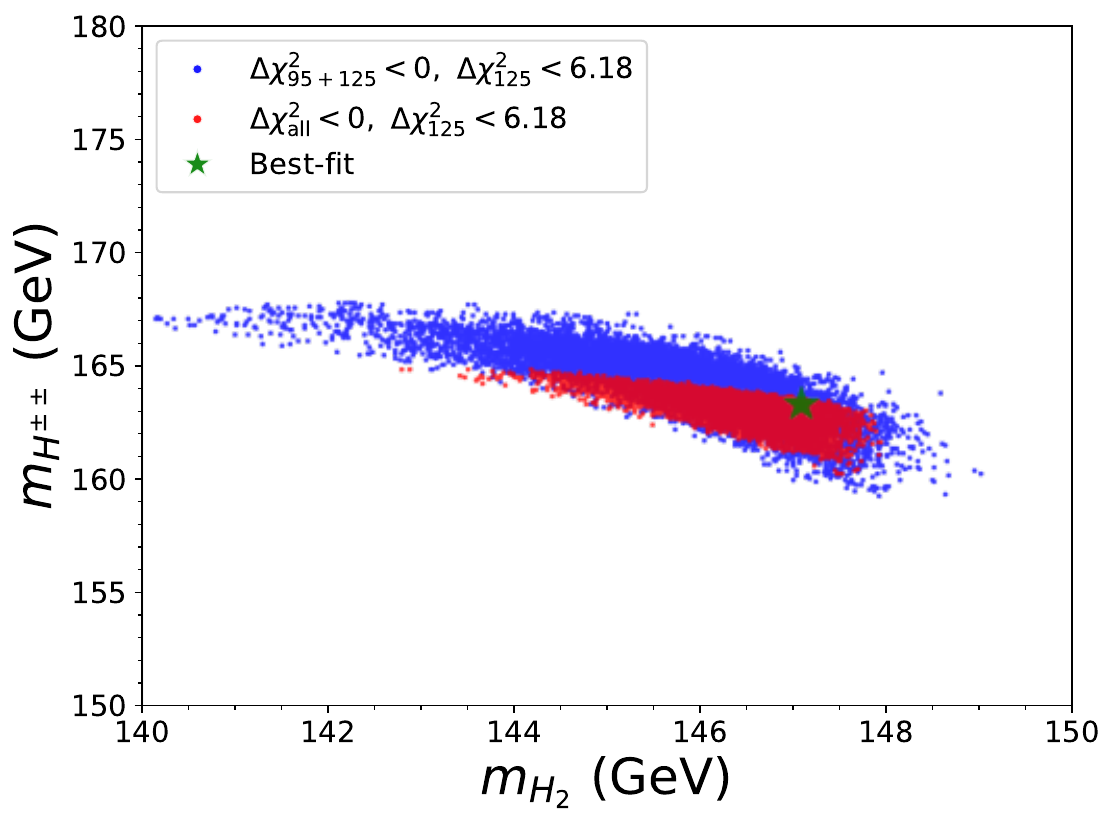}
    \includegraphics[width=0.48\textwidth]{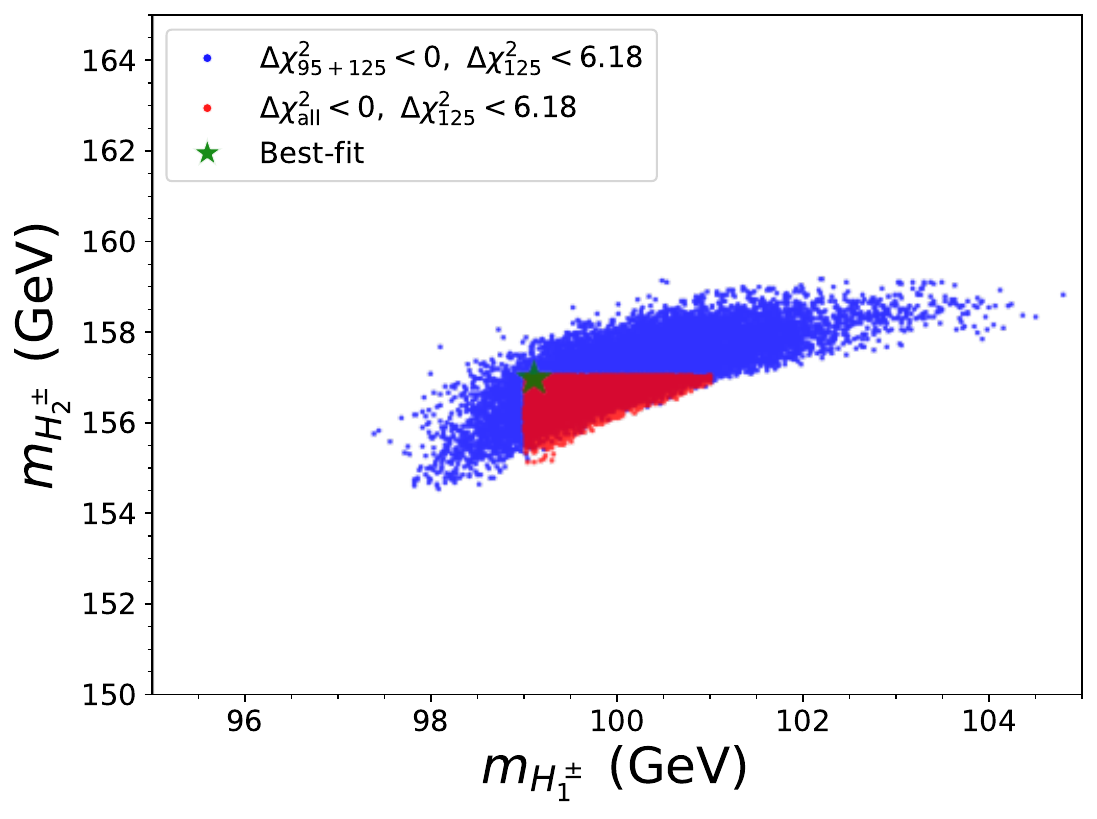}
    \caption{Sample distributions in the ({\it left}) $m_{H_2}$--$m_{H^{\pm\pm}}$ and ({\it right}) $m_{H_1^\pm}$--$m_{H_2^{\pm}}$ planes. The color coding is the same as in \protect\reffi{fig:vev-mu_H1}.
    }
    \label{fig:masses}
\end{figure}

In \reffi{fig:masses}, we show the sample distributions in the $m_{H_2}$--$m_{H^{\pm\pm}}$ (left) and $m_{H_1^{\pm}}$--$m_{H_2^{\pm}}$ (right) planes, with the same color coding as in \reffi{fig:vev-mu_H1}. One can observe that all scalar masses are $\sim\mathcal{O}(100)$~GeV, where the sharp cuts on the red-point distributions are caused by the requirement of $m_{H_1^\pm}\in[99,101]$~GeV and  $m_{H_2^\pm}\in[155,157]$~GeV. Consequently, the mass hierarchy of the \megm model preferred by the considered constraints is
\begin{equation}
    m_{H_1}=95 \gev < m_{H_1^\pm} < m_h=125 \gev < m_{H_2} < m_{H_3}=152 \gev < m_{H_2^\pm} < m_{H^{\pm\pm}} ~.
\end{equation}
The reason giving rise to the light scalar masses in the meGM model presented here is similar to that explained in Ref.~\cite{Chen:2023bqr} for the GM model. In the GM model, if the SM-like Higgs boson is approximately decoupled from the other scalar fields, the other scalar masses will follow the approximate relation
\begin{equation}
    2m_{H_{\rm 1,GM}}^2 \approx 
    3m_{H_{\rm 3,GM}}^2 - m_{H_{\rm 5,GM}}^2 ~,
\end{equation}
where $m_{H_{\rm i,GM}}$ for $i=1,3,5$ are the exotic singlet, triplet, and quintet scalar masses. Once we fix the value of two of the masses (e.g., $m_{H_{\rm 1,GM}}=95$~GeV and $m_{H_{\rm 5,GM}}=152$~GeV), the remaining mass will then also be of roughly the same scale. Since the meGM model is constructed in a way that it only violates the custodial symmetry mildly and that the 125-GeV scalar is found to be SM-like according to the LHC measurements (and imposed through \texttt{HiggsSignals} in our numerical analysis), we find the same outcome of similarly-scaled scalar masses once we fix $m_{H_1}=95$~GeV and $m_{H_5}=152$~GeV.

%%%%%%%%%%%%%%%%%%%%%%%%%%% F I G U R E %%%%%%%%%%%%%%%%%%%%%%%%%%%%%%%%%%%%%%%%%%%%%%%%%%%%%%%%%%%%%%%%%%
%%%%%%%%%%%%%%%%%%%%%%%%%%% F I G U R E %%%%%%%%%%%%%%%%%%%%%%%%%%%%%%%%%%%%%%%%%%%%%%%%%%%%%%%%%%%%%%%%%%

\begin{figure}[ht!]
    \centering
    \includegraphics[width=0.48\textwidth]{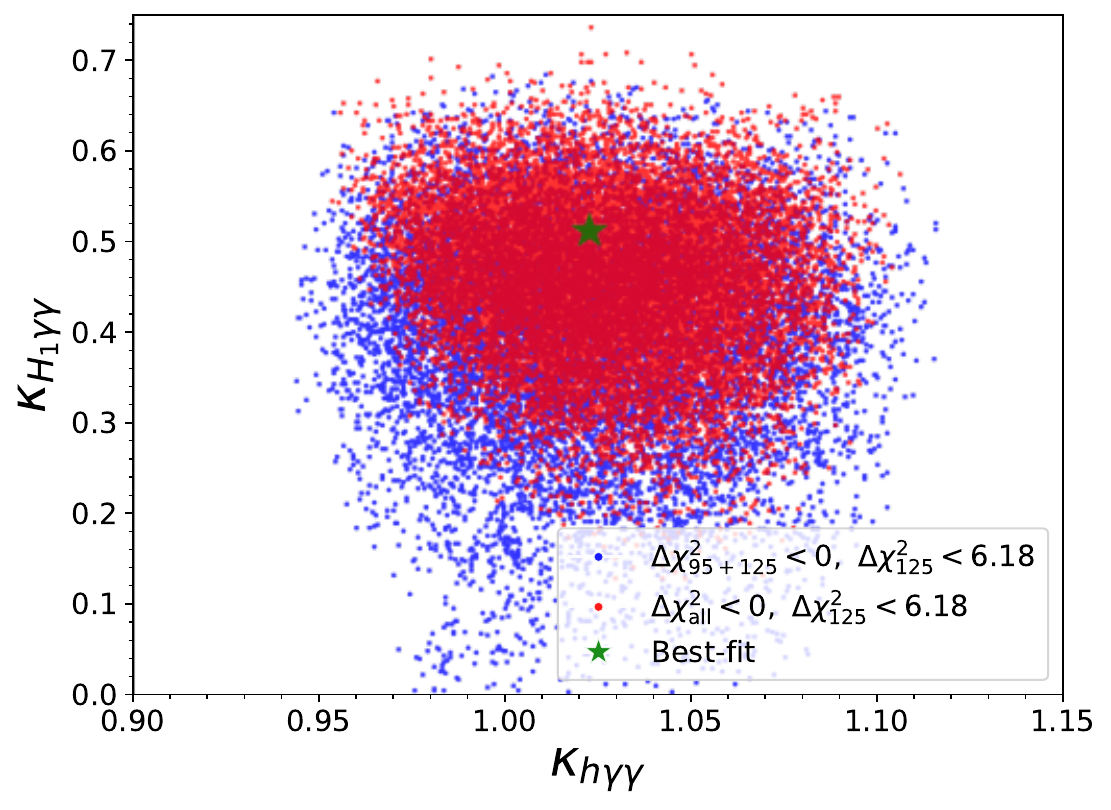}
    \includegraphics[width=0.48\textwidth]{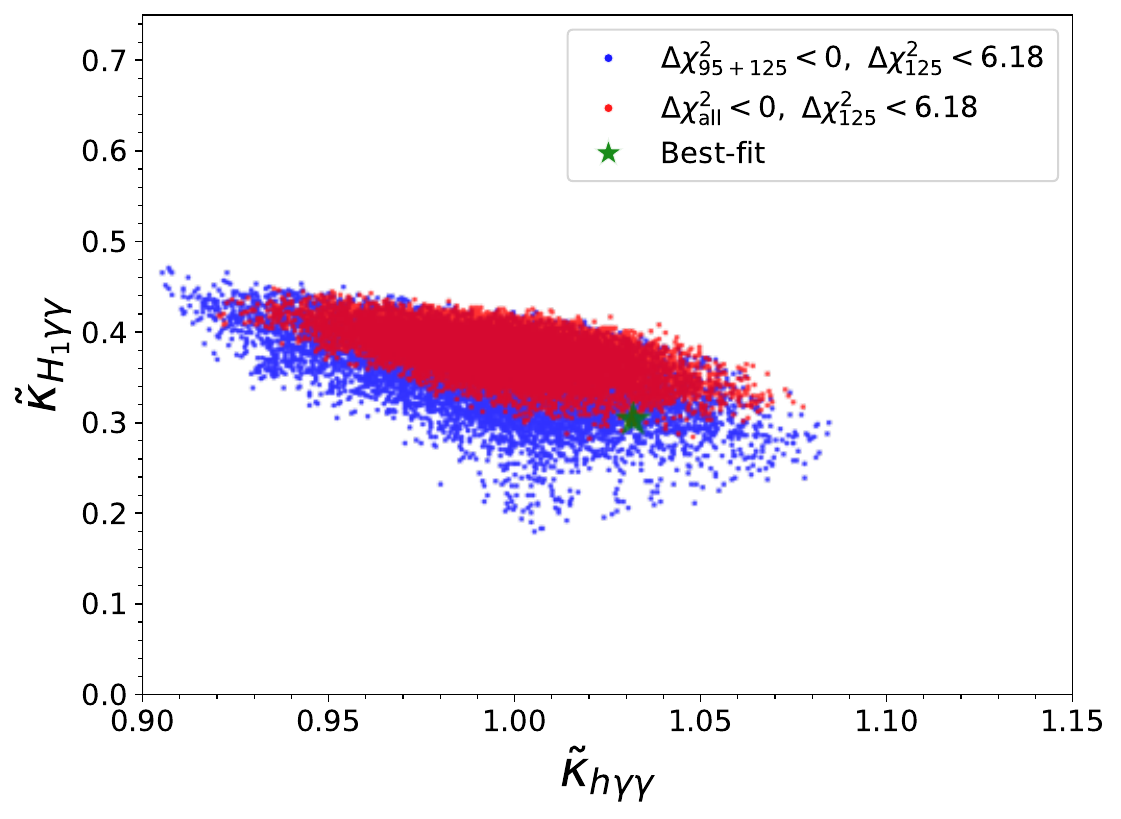}
    \caption{Sample distributions in the ({\it left}) $\kappa_{h\gamma\gamma}$--$\kappa_{H_1\gamma\gamma}$ and ({\it right}) $\tilde{\kappa}_{h\gamma\gamma}$--$\tilde{\kappa}_{H_1\gamma\gamma}$ planes (see text for details). The color coding is the same as in \protect\reffi{fig:vev-mu_H1}.
    }
    \label{fig:khgaga-kH1gaga}
\end{figure}

In \reffi{fig:khgaga-kH1gaga}, we compare the results for $h \to \gamma\gamma$ and $H_1 \to \gamma\gamma$, including (left plot) or excluding (right plot) the contribution of the $H^{\pm\pm}$~loop. From the distribution in the $\kappa_{h\gamma\gamma}$--$\kappa_{H_1\gamma\gamma}$ plane shown in the left panel one can see that the obtained $\kappa_{h\gamma\gamma}$ values lie in the interval $[0.95,1.12]$, which is consistent with the SM predictions as reflected by the Higgs measurements to date, and the preferred range for $\kappa_{H_1\gamma\gamma}$ is roughly $[0.2, 0.7]$. While as defined above $\Gamma_{H_i\to\gamma\gamma}^{\rm eGM}$ entering $\kappa_{H_i\gamma\gamma}$ contains all contributing particles in the loop, for illustrating the effects of the $H^{\pm\pm}$-loop contributions, we furthermore define
\begin{equation}
   \tilde{\kappa}_{H_i\gamma\gamma} = \left(\frac{\tilde{\Gamma}_{H_i\to\gamma\gamma}^{\rm eGM}}{\Gamma_{H_i\to\gamma\gamma}^{\rm SM}}\right)^{1/2} ~.
\end{equation}
Here $\tilde{\Gamma}_{H_i\to\gamma\gamma}^{\rm eGM}$ denotes the one-loop di-photon decay width of $H_i$ where the $H^{\pm\pm}$-loop contribution predicted by the eGM model is left out. By comparing the two panels, one can see that the $H^{\pm\pm}$ contribution plays a major role in enhancing the di-photon couplings of $H_1$ except for some of the blue points, clearly demonstrating the significance of this feature of the (e)GM-type models. The di-photon coupling of the Higgs boson at $125 \gev$, on the other hand, is less affected by the $H^{\pm\pm}$-loop contribution and remains in the experimentally preferred range for both displayed cases.

%%%%%%%%%%%%%%%%%%%%%%%%%%% F I G U R E %%%%%%%%%%%%%%%%%%%%%%%%%%%%%%%%%%%%%%%%%%%%%%%%%%%%%%%%%%%%%%%%%%
%%%%%%%%%%%%%%%%%%%%%%%%%%% F I G U R E %%%%%%%%%%%%%%%%%%%%%%%%%%%%%%%%%%%%%%%%%%%%%%%%%%%%%%%%%%%%%%%%%%

%%%%%%%%%%%%%%%%%%%%%%%%%%% F I G U R E %%%%%%%%%%%%%%%%%%%%%%%%%%%%%%%%%%%%%%%%%%%%%%%%%%%%%%%%%%%%%%%%%%
\begin{figure}[ht!]
    \centering
    \includegraphics[width=0.48\textwidth]{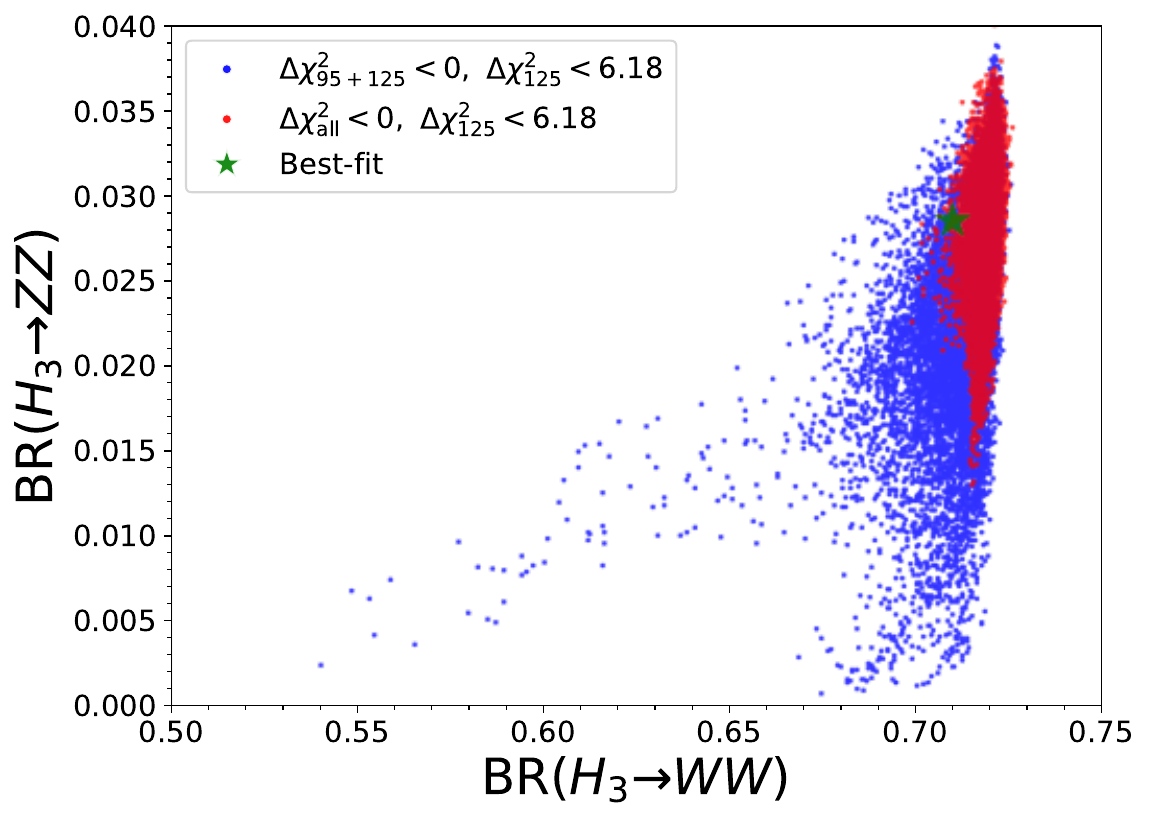}
    \includegraphics[width=0.48\textwidth]{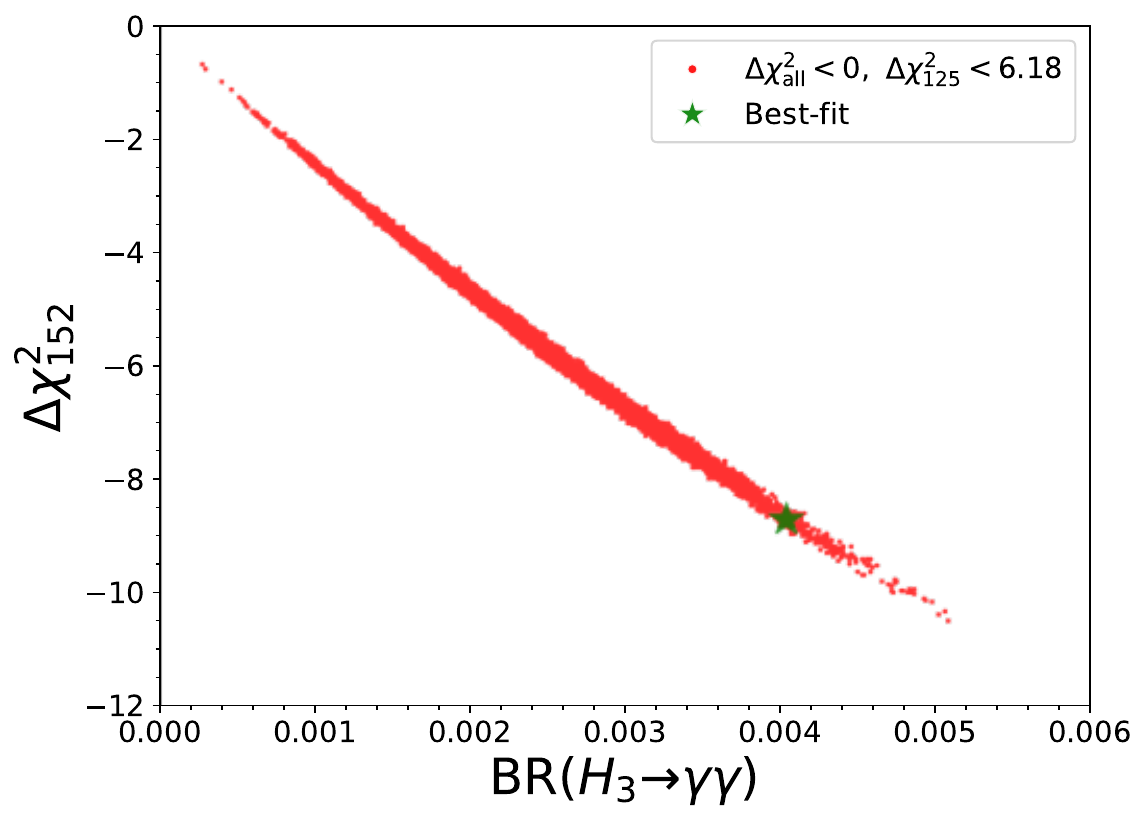}
    \caption{Sample distributions in the ({\it left}) ${\rm BR}(H_3\to WW)$--${\rm BR}(H_3\to ZZ)$ and ({\it right}) ${\rm BR}(H_3\to \gamma\gamma)$--$\Delta\chi^2_{152}$ planes. The color coding is the same as in \protect\reffi{fig:vev-mu_H1}. It should be noted that we did not include the blue points in the right plot since we do not perform the 152-GeV excess analysis on them.
    }
    \label{fig:BRH3}
\end{figure}
%%%%%%%%%%%%%%%%%%%%%%%%%%% F I G U R E %%%%%%%%%%%%%%%%%%%%%%%%%%%%%%%%%%%%%%%%%%%%%%%%%%%%%%%%%%%%%%%%%%

Next, we show in \reffi{fig:BRH3} the sample distributions in the ${\rm BR}(H_3\to WW)$--${\rm BR}(H_3\to ZZ)$ (left) and ${\rm BR}(H_3\to \gamma\gamma)$--$\Delta\chi^2_{152}$ (right) planes. It should be noted that we do not perform the 152-GeV excess analysis on the blue points, and hence we do not show them in the right plot. One can see that ${\rm BR}(H_3\to WW)$ is for most points more than an order of magnitude larger than ${\rm BR}(H_3\to ZZ)$, which is caused by both the breaking of the custodial symmetry (and thus the deviation from the originally fixed ratio between the $H_3WW$ and $H_3ZZ$ couplings in the GM model) and the different kinematic suppression associated with the $2W$ and $2Z$ mass thresholds. On the other hand, ${\rm BR}(H_3\to \gamma\gamma)$ is almost linearly proportional to $\Delta\chi^2_{152}$ with a negative slope, and values of $\Delta\chi^2_{152} \le -10$ can be reached. Thus, a good description of the data on the excess at about 152~GeV is correlated with relatively large values of ${\rm BR}(H_3\to \gamma\gamma)$.

\begin{figure}[htb!]
    \centering
    \includegraphics[width=0.24\linewidth]{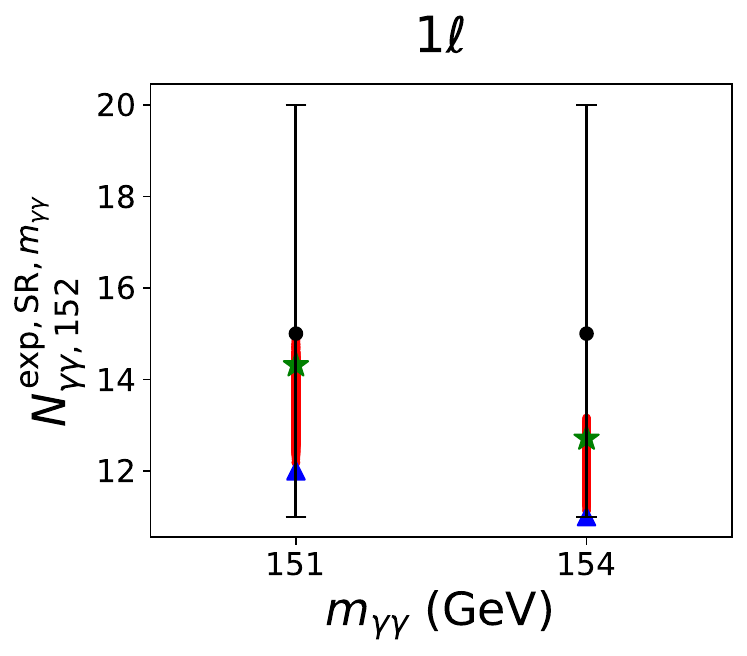}
    \includegraphics[width=0.24\linewidth]{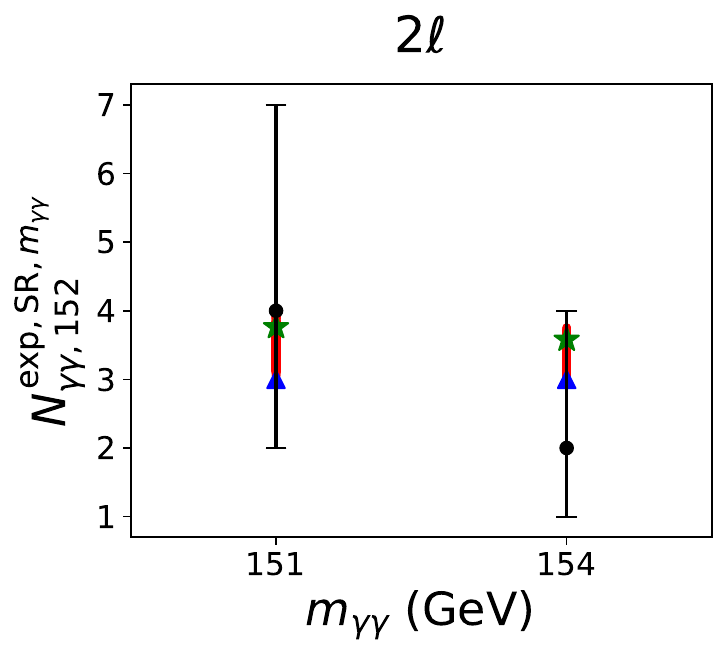}
    \includegraphics[width=0.24\linewidth]{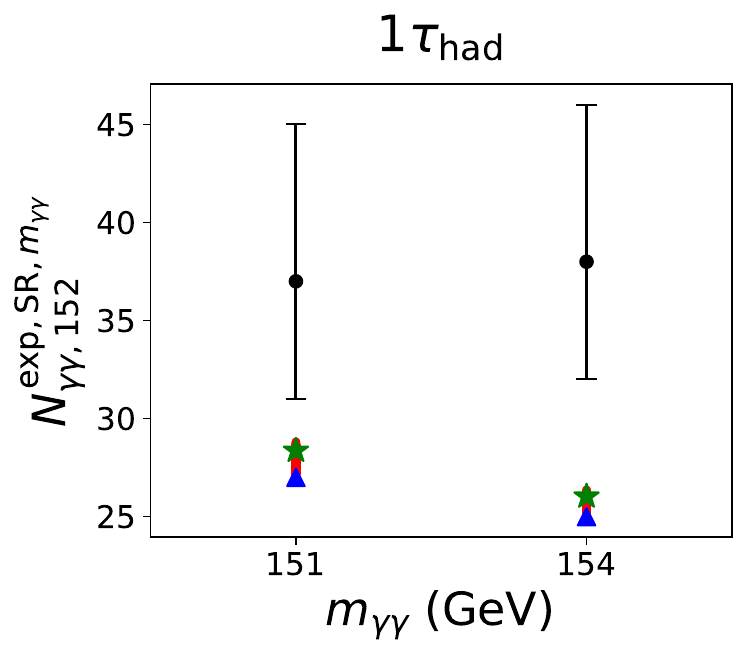}
    \includegraphics[width=0.24\linewidth]{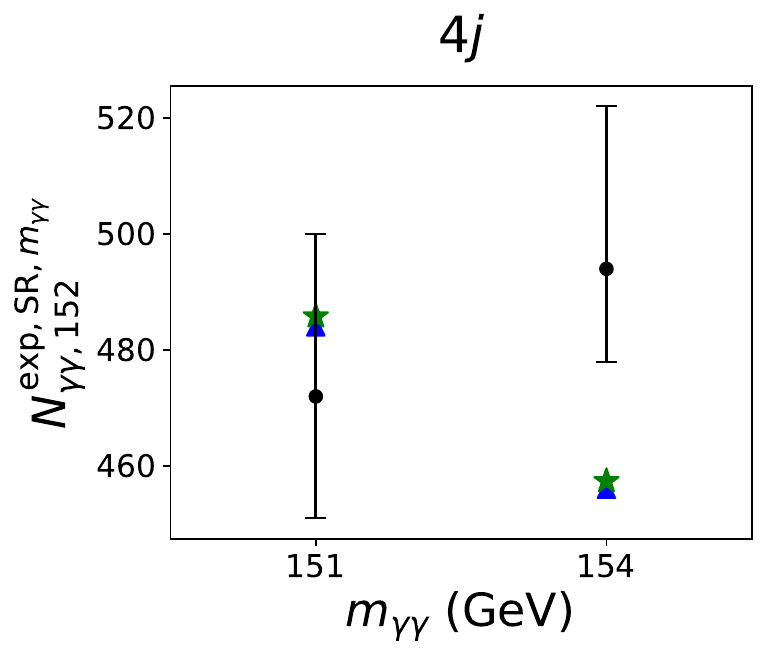}
    \includegraphics[width=0.24\linewidth]{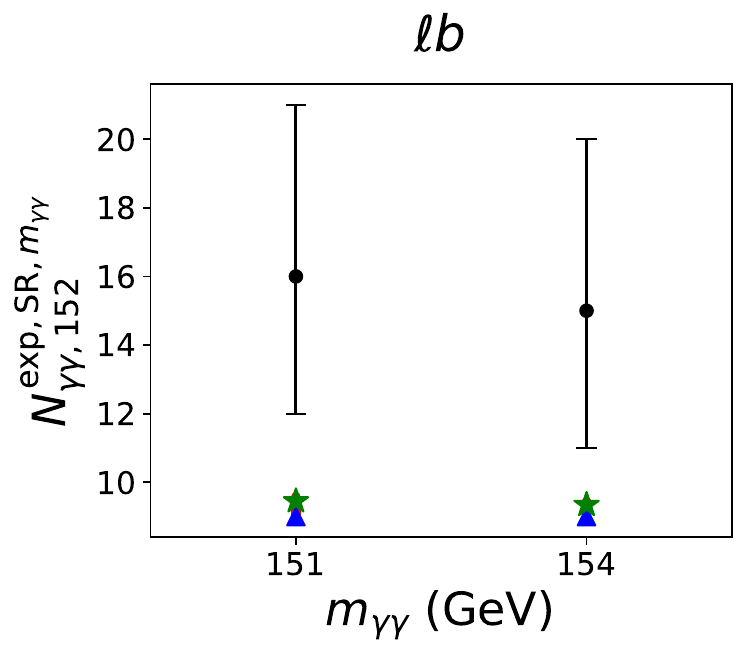}
    \includegraphics[width=0.24\linewidth]{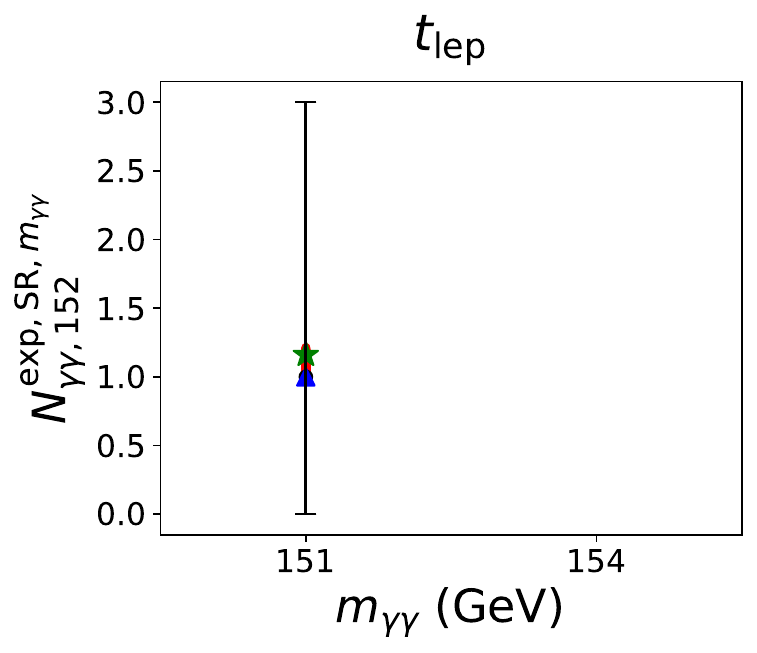}
    \includegraphics[width=0.24\linewidth]{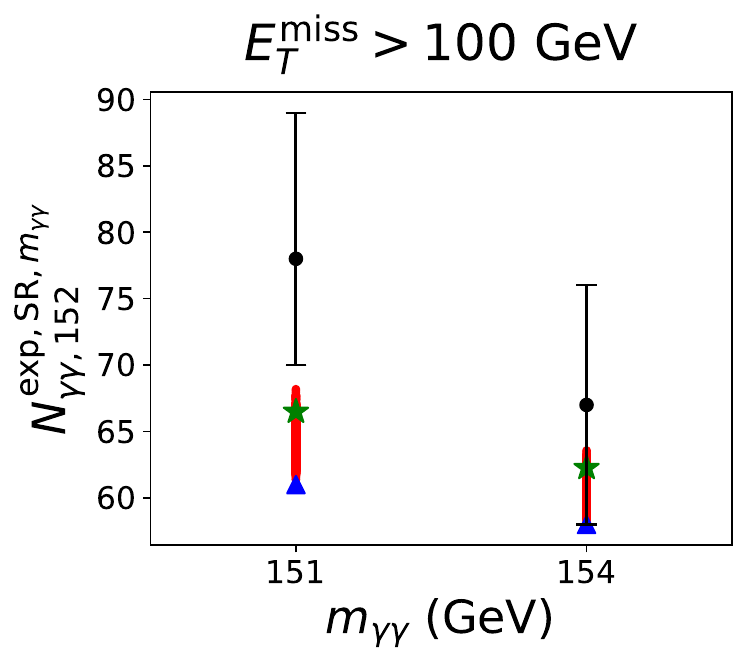}
    \includegraphics[width=0.24\linewidth]{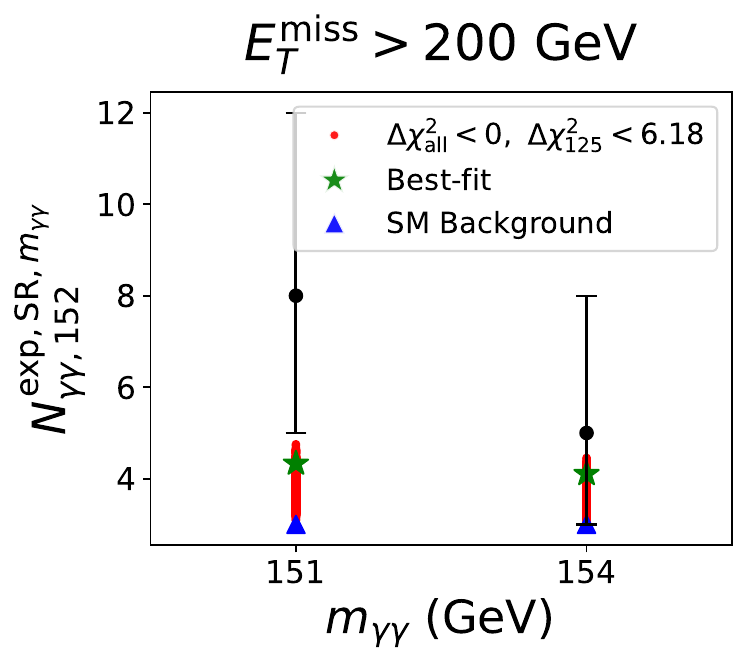}
    \caption{Sample distributions in the eight relevant SR's listed in Tab.~\ref{tab:SR} in the $m_{\gamma\gamma}$--$N_{\gamma\gamma,152}^{{\rm exp, SR},m_{\gamma\gamma}}$ plane for $m_{\gamma\gamma}=151,154$~GeV. The color coding in each panel is the same as in \protect\reffi{fig:vev-mu_H1}. The data from Refs.~\cite{ATLAS:2023omk,ATLAS:2024lhu} are shown in terms of the black dots and error bars. 
    The best-fit point is indicated by a green star, while the SM background fitted by ATLAS~\cite{ATLAS:2023omk,ATLAS:2024lhu} 
    is shown as a blue triangle. We note that in certain panels where the red dots seem invisible, they are in fact covered by the green star and the blue triangle.
    }
    \label{fig:H152_SRs}
\end{figure}

In order to investigate the compatibility of the \megm\ predictions with the data in the individual SR's, we show in \reffi{fig:H152_SRs} the red points of \reffis{fig:vev-mu_H1}--\ref{fig:BRH3} in the SR's listed in \refta{tab:SR} in the $m_{\gamma\gamma}$--$N_{\gamma\gamma,152}^{{\rm exp, SR},m_{\gamma\gamma}}$ plane for $m_{\gamma\gamma}=151,154 \gev$. We also show the best-fit point (green star) and the SM background (blue triangle) fitted by ATLAS (see Refs.~\cite{ATLAS:2023omk,ATLAS:2024lhu}). It should be noted that in certain panels where the red dots seem invisible, they are in fact covered by the green star and the blue triangle. We furthermore note that there is no measurement at $154$~GeV in the $t_{\rm lep}$ SR (second to the left plot in the bottom row in \reffi{fig:H152_SRs}). One can see that the \megm\ model can describe the excesses in the $1\ell$ SR at both bins, $2\ell$ SR at $151$~GeV, and the $E_T^{\rm miss}>100,200$~GeV SR's at $154$~GeV much better than the SM. On the other hand, the \megm\ model adds only marginally more flexibility for describing the other bins (or even leads to slightly worse fits there than the SM).

%%%%%%%%%%%%%%%%%%%%%%%%%%% F I G U R E %%%%%%%%%%%%%%%%%%%%%%%%%%%%%%%%%%%%%%%%%%%%%%%%%%%%%%%%%%%%%%%%%%
%%%%%%%%%%%%%%%%%%%%%%%%%%% F I G U R E %%%%%%%%%%%%%%%%%%%%%%%%%%%%%%%%%%%%%%%%%%%%%%%%%%%%%%%%%%%%%%%%%%

The physical properties predicted for the best-fit point are summarized in \refta{tab:best-fit}, which indicates the phenomenologically preferred parameters and decay channels in that region of the parameter space around the best-fit point. We first note that the overall $\Delta\chi^2=-19.2$ corresponds to a better description of the data than for the SM, which also holds individually for the excesses at 95~GeV, $\Delta\chi^2_{95}=-11.0$, and 152~GeV, $\Delta\chi^2_{152}=-8.72$. Next, one can see that the mass spectrum does not show degeneracy, which is consistent with the custodial-symmetry violating hierarchy of ${\rm BR}(H_3\to WW)\gg {\rm BR}(H_3\to ZZ)$. Moreover, in accordance with the general numerical simulations presented earlier, the dominant decay channel of the heavier $H_2^\pm$ is to $H_1^\pm Z$ with a BR of 0.764, followed by the $WZ$ mode with a BR of 0.111. Also the neutral state $H_2$ decays predominantly to other light scalars, with ${\rm BR}(H_2\to H_1Z)=0.772$ and ${\rm BR}(H_2\to H_1^\pm W^\mp)=0.127$, while $H^{\pm\pm}$ also decays to $H_1^\pm W^\pm$ with a relatively large BR of 0.297. Accordingly, the description of the 95-GeV and 152-GeV excesses in the meGM model implies a rich phenomenology among the different scalar bosons which can be probed in future experiments.

%%%%%%%%%%%%%%%%%%%%%%%%%%%%%%%%%% T A B L E %%%%%%%%%%%%%%%%%%%%%%%%%%%%%%%%%%%%%%%%%%%%%%%%%%%%%%%%%%%
\begin{table}[ht!]
    \centering
    \begin{tabular}{|c||c|c|c|c|c|c|c|c|}
    \toprule
    \multicolumn{9}{|c|}{Best-fit Point Properties} \\
    \toprule
    \multicolumn{9}{|c|}{~~$v_\chi=4.60$~GeV,~$v_\xi=5.01$~GeV,~$m_{H_2}=147$~GeV~~} \\
    \multicolumn{9}{|c|}{~~$m_{H_1^\pm}=100$~GeV,~$m_{H_2^\pm}=156$~GeV,~$m_{H^{\pm\pm}}=163$~GeV~~} \\
    \multicolumn{9}{|c|}{~~$\kappa_{hWW}=0.940$,~$\kappa_{hZZ}=0.969$,~$\kappa_{hff}=0.957$~~} \\
    \multicolumn{9}{|c|}{~~$\kappa_{H_1WW}=0.299$,~$\kappa_{H_1ZZ}=0.239$,~$\kappa_{H_1ff}=0.208$~~} \\
    \multicolumn{9}{|c|}{~~${\rm BR}(H_3\to\gamma\gamma)=0.004$,~$\Delta\chi^2=-19.2$,~$\Delta\chi^2_{95}=-11.0$,~$\Delta\chi^2_{152}=-8.72$~~} \\
    \colrule
    \multirow{2}{*}{$\mu_{95,XX}$} & \multicolumn{4}{c|}{$b\bar{b}$} & \multicolumn{4}{c|}{$\gamma\gamma$} \\
    \cline{2-9}
    & \multicolumn{4}{c|}{0.058} & \multicolumn{4}{c|}{0.265} \\
    \colrule
    \multirow{2}{*}{~~${\rm BR}(h\to XX)$~~} & $b\bar{b}$ & $cc$ & $\tau\tau$ & $gg$ & $\gamma\gamma$ & $Z\gamma$ & $ZZ$ & $WW$ \\
    \cline{2-9}
    & ~~0.578~~ & ~~0.029~~ & ~~0.062~~ & ~~0.081~~ & ~~0.0026~~ & ~~0.002~~ & ~0.027~ & ~~0.207~~ \\
    \colrule
    \multirow{2}{*}{~${\rm BR}(H_1\to XX)$~} & $b\bar{b}$ & $cc$ & $\tau\tau$ & $gg$ & $\gamma\gamma$ & $Z\gamma$ & $ZZ$ & $WW$ \\
    \cline{2-9}
    & ~0.822~ & ~0.041~ & ~0.084~ & ~0.034~ & ~0.008~ & ~0.000~ & ~0.001~ & ~0.010~ \\
    \colrule
    \multirow{2}{*}{~${\rm BR}(H_3\to XX)$~} & $b\bar{b}$ & $cc$ & $\tau\tau$ & $gg$ & $Z\gamma$ & $ZZ$ & $WW$ & $H_1^\pm W$ \\
    \cline{2-9}
    & ~0.171~ & ~0.008~ & ~0.019~ & ~0.023~ & ~0.003~ & ~0.028~ & ~0.710~ & ~0.033~ \\
    \colrule
    \multirow{2}{*}{~${\rm BR}(H_2\to XX)$~} & $b\bar{b}$ & $cc$ & $\tau\tau$ & $gg$ & $\gamma\gamma$ & $hZ$ & $H_1Z$ & $H_1^\pm W$ \\
    \cline{2-9}
    & ~0.066~ & ~0.003~ & ~0.007~ & ~0.022~ & ~0.000~ & ~0.002~ & ~0.772~ & ~0.127~ \\
    \colrule
    \multirow{2}{*}{~${\rm BR}(H_1^\pm\to XX)$~} & \multicolumn{4}{c|}{$cs$} & \multicolumn{4}{c|}{$\tau\nu_\tau$} \\
    \cline{2-9}
    & \multicolumn{4}{c|}{0.612} & \multicolumn{4}{c|}{0.379} \\
    \colrule
    \multirow{2}{*}{~${\rm BR}(H_2^\pm\to XX)$~} & $cs$ & $\tau\nu_\tau$ & $tb$ & $WZ$ & $H_1^\pm Z$ & $hW$ & $H_1W$ & $H_2W$ \\
    \cline{2-9}
    & ~0.006~ & ~0.004~ & ~0.023~ & ~0.111~ & ~0.764~ & ~0.029~ & ~0.060~ & ~0.003~ \\
    \colrule
    \multirow{2}{*}{~${\rm BR}(H^{\pm\pm}\to XX)$~} & \multicolumn{4}{c|}{$WW$} & \multicolumn{4}{c|}{$H_1^\pm W$} \\
    \cline{2-9}
    & \multicolumn{4}{c|}{0.702} & \multicolumn{4}{c|}{0.297} \\
    \botrule
    \end{tabular}
    \vspace{1em}
    \caption{Summary of the physical properties predicted for the best-fit point.}
    \label{tab:best-fit}
    \vspace{2em}
\end{table}
%%%%%%%%%%%%%%%%%%%%%%%%%%%%%%%%%% T A B L E %%%%%%%%%%%%%%%%%%%%%%%%%%%%%%%%%%%%%%%%%%%%%%%%%%%%%%%%%%%

Finally, we comment on the comparison of the SM and the \megm\ model with respect to the region of $m_{H_3} \approx 152 \gev$. The SM has $\chi^2_{152}({\rm SM}) \approx 43.1$ for 37 bins, reflecting the (small) tension of the experimental data with the SM prediction. The \megm\ model can improve this to $\chi^2_{152,{\rm min}}({\rm \megm}) \approx  33$ for the same number of bins. While this improvement appears to be moderate, it reduces the $\chi^2_{152}$ value by about 23\% and relaxes the tension in comparison to the SM. Future experimental data will be decisive to either marginalize or substantiate this (small) tension in the SM and the possible improvement within the \megm\ model.

%================================================================
\section{Future prospects}\label{sec:prospects}

\begin{figure}[ht!]
    \centering
    \includegraphics[width=0.47\textwidth]{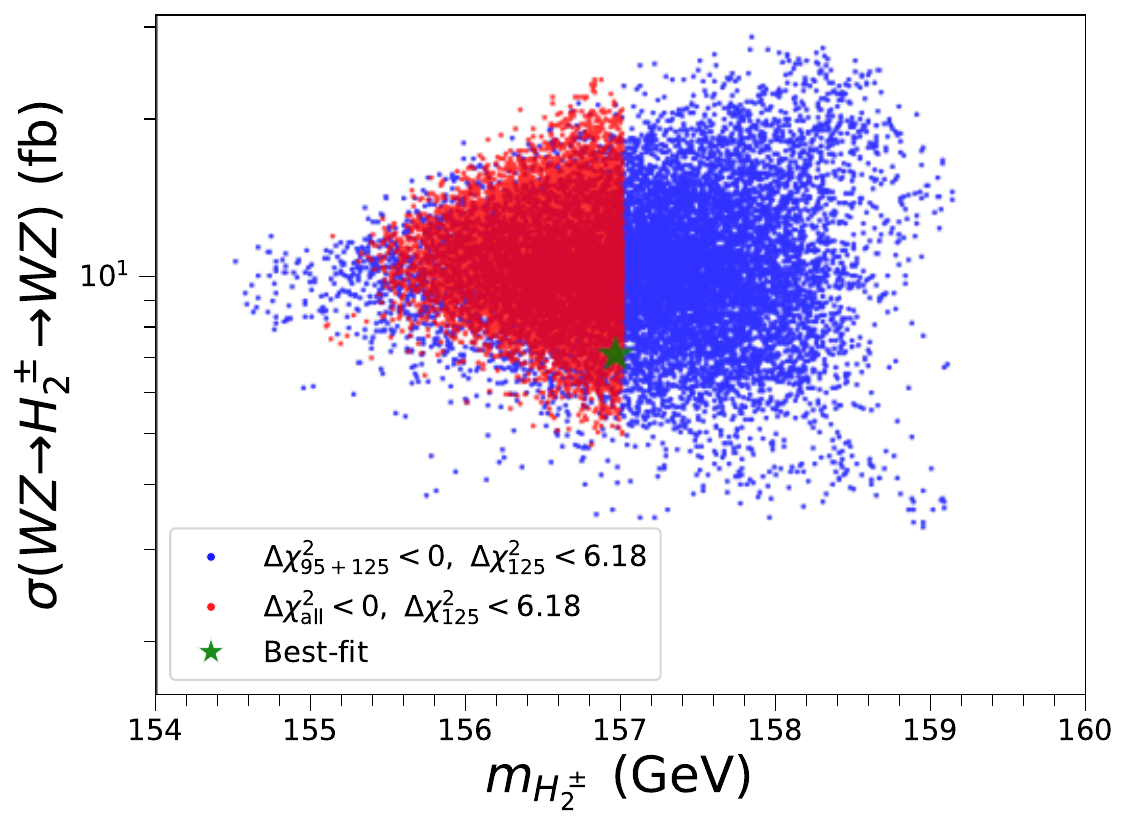}
    \includegraphics[width=0.49\textwidth]{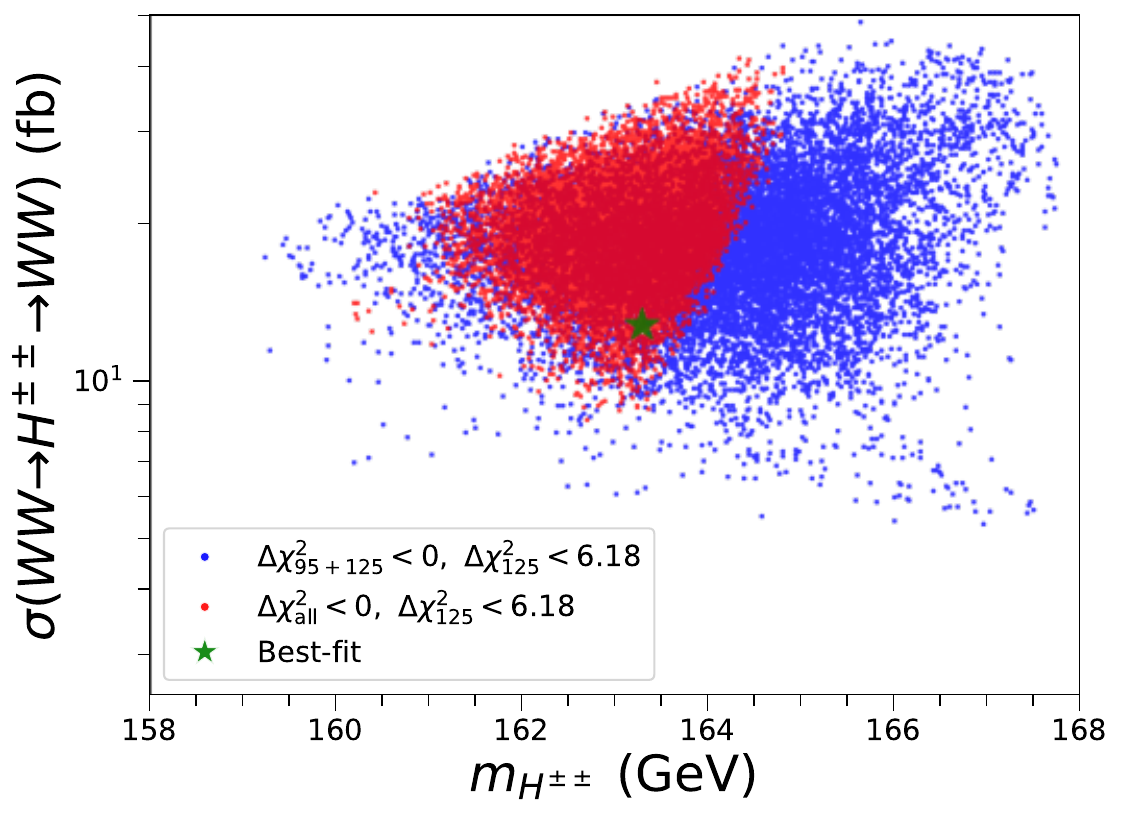}
    \caption{Sample distributions in the ({\it left}) $m_{H_2}^\pm$--$\sigma(WZ\to H_2^\pm\to WZ)$ and ({\it right}) $m_{H^{\pm\pm}}$--$\sigma(WW\to H^{\pm\pm}\to WW)$ planes at the 14-TeV (HL-)LHC. The color coding is the same as in \protect\reffi{fig:vev-mu_H1}.
    }
    \label{fig:sensitivity}
\end{figure}

We finish our analysis by investigating the prospects for testing the \megm\ scenarios by providing simultaneously a description of the excess at 95~GeV together with the signal at 125~GeV and an interpretation of a possible excess around $152 \gev$. We start with an analysis of the most promising search channels at the HL-LHC in \reffi{fig:sensitivity}, in which we show the two samples in the $m_{H_2}^\pm$--$\sigma(WZ\to H_2^\pm\to WZ)$ (left) and $m_{H^{\pm\pm}}$--$\sigma(WW\to H^{\pm\pm}\to WW)$ (right) planes at the 14-TeV (HL-)LHC. Compared to the similar $WZ\to H_5^\pm\to WZ$ and $WW\to H_5^{\pm\pm}\to WW$ channels in the GM model as analyzed in \citere{Chen:2023bqr}, the cross sections predicted by the \megm\ model presented here both tend to be somewhat enhanced, making these channels more promising to probe the scenario at the (HL-)LHC.

%%%%%%%%%%%%%%%%%%%%%%%%%%% F I G U R E %%%%%%%%%%%%%%%%%%%%%%%%%%%%%%%%%%%%%%%%%%%%%%%%%%%%%%%%%%%%%%%%%%
%%%%%%%%%%%%%%%%%%%%%%%%%%% F I G U R E %%%%%%%%%%%%%%%%%%%%%%%%%%%%%%%%%%%%%%%%%%%%%%%%%%%%%%%%%%%%%%%%%%

\begin{figure}[ht!]
    \centering
    \includegraphics[width=0.75\textwidth]{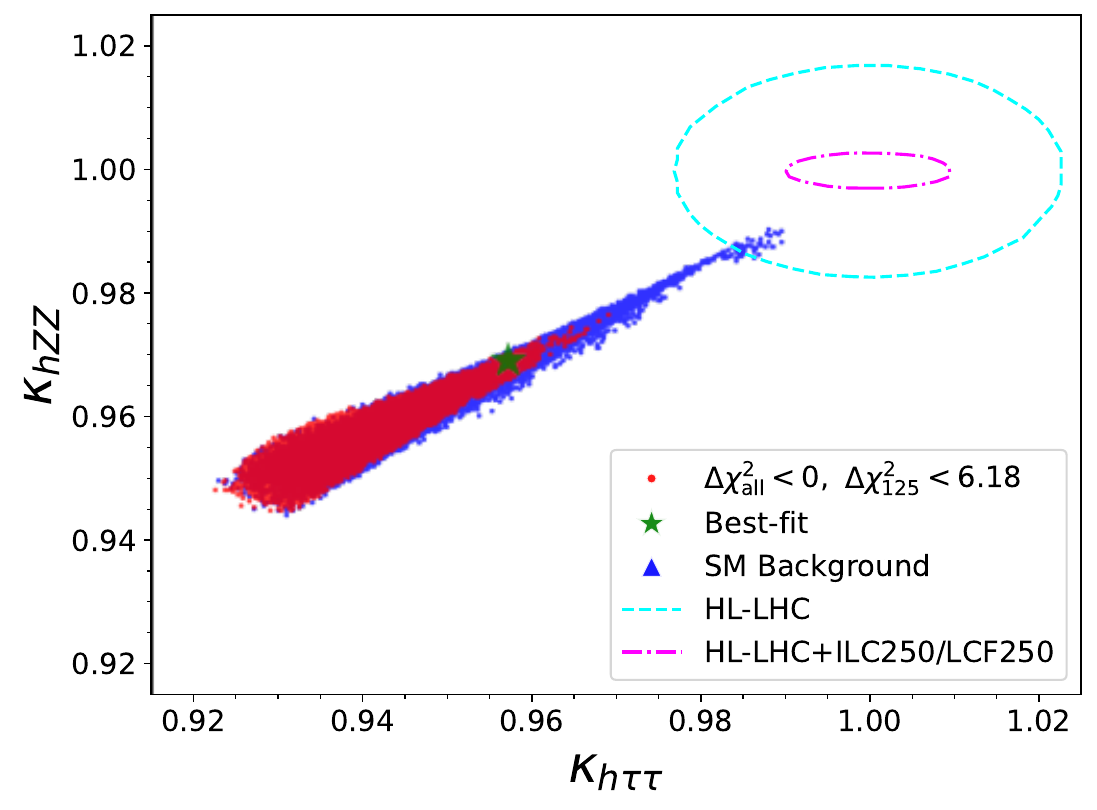}
    \caption{Sample distribution in the $\kappa_{h\tau\tau}$--$\kappa_{hZZ}$ plane in comparison to the prospective precision 
    at the $1\,\sigma$ level (indicated for the SM value) at the HL-LHC (cyan)~\cite{deBlas:2025gyz,HL-LHC-input} and the 
    HL-LHC+ILC250/LCF250 (magenta)~\cite{Bambade:2019fyw,deBlas:2025gyz,LCF-input}. 
    The other color coding is the same as in \protect\reffi{fig:vev-mu_H1}
    }
    \label{fig:khff_vs_khZZ}
\end{figure}

Next, we analyze the potential of future $e^+e^-$ colliders (such as LCF, ILC, CLIC, FCC-ee, CEPC) to further probe the considered \megm\ 
scenarios. We first present in \reffi{fig:khff_vs_khZZ} the two samples in the $\kappa_{h\tau\tau}$--$\kappa_{hZZ}$ plane 
(i.e., the coupling modifiers of the $125 \gev$ Higgs boson with respect to the SM prediction) overlaid with the anticipated precisions
for the coupling measurements at the HL-LHC (cyan dashed)~\cite{deBlas:2025gyz,HL-LHC-input} and combined with (hypothetical future) 
ILC250/LCF250 measurement accuracies (magenta dot-dashed)~\cite{Bambade:2019fyw,deBlas:2025gyz,LCF-input} (denoted
as ``LCF250'' in the plot).
The $1\,\sigma$ contours in the plot are centered around the SM prediction of $\kappa_{h\tau\tau} = \kappa_{hZZ} = 1$. 
The deviations predicted in the couplings of the Higgs boson at 125~GeV with respect to the SM predictions are seen to be substantial in 
comparison to the projected accuracies, while, as explained above, they are compatible with the present signal rate measurements 
at the LHC (as tested with \texttt{HiggsSignals}). In comparison to the coupling values predicted by the SM, nearly all of the blue 
and all of the red sample points are outside the $1\,\sigma$ HL-LHC ellipse, and the best-fit point shows a $\sim 3\,\sigma$ deviation. The sensitivity is substantially improved for the case where the prospective ILC250/LCF250 measurements are included. 
All points in the preferred sample (red) would yield a deviation of more than $5\,\sigma$.

%%%%%%%%%%%%%%%%%%%%%%%%%%% F I G U R E %%%%%%%%%%%%%%%%%%%%%%%%%%%%%%%%%%%%%%%%%%%%%%%%%%%%%%%%%%%%%%%%%%
%%%%%%%%%%%%%%%%%%%%%%%%%%% F I G U R E %%%%%%%%%%%%%%%%%%%%%%%%%%%%%%%%%%%%%%%%%%%%%%%%%%%%%%%%%%%%%%%%%%

\begin{figure}[ht!]
    \centering
    \includegraphics[width=0.75\textwidth]{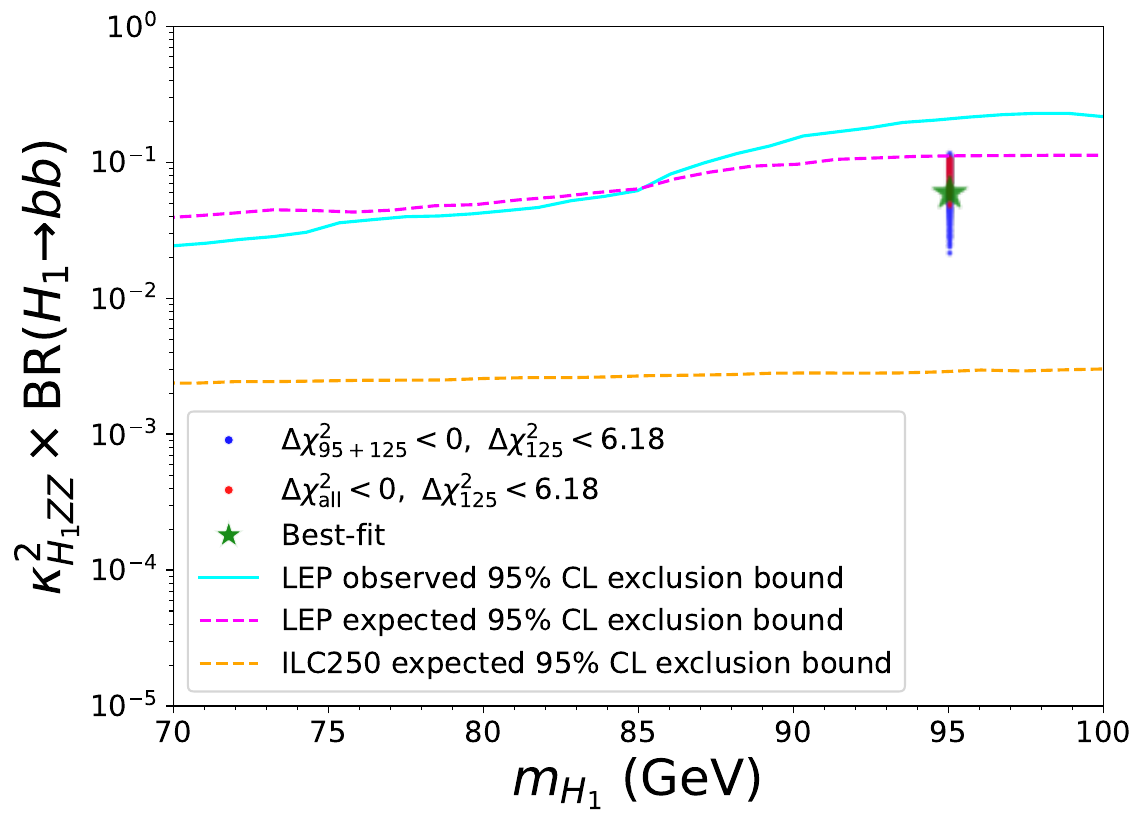}
    \caption{Sample distribution in the $m_{H_1}$--$\kappa_{H_1ZZ}^2\times{\rm BR}(H_1\to b\bar{b})$ plane in comparison to the 95\% CL LEP observed exclusion bound~\cite{LEPWorkingGroupforHiggsbosonsearches:2003ing} (cyan), the 95\% CL LEP expected exclusion bound~\cite{LEPWorkingGroupforHiggsbosonsearches:2003ing} (magenta), and a projection for the 95\% CL ILC250 expected exclusion bound~\cite{Drechsel:2018mgd} (orange). The other color coding is the same as in \protect\reffi{fig:vev-mu_H1}.
    }
    \label{fig:mH1_Lepton}
\end{figure}

Finally, we analyze the capability of the ILC to produce the new Higgs boson at $\sim 95$~GeV, \ie, $H_1$, and to measure its couplings. In \reffi{fig:mH1_Lepton}, we show the $m_{H_1}$--$\kappa_{H_1ZZ} \times {\rm BR}(H_1 \to b\bar b)$ plane. In the plot, the cyan (magenta) curve indicates the observed (expected) exclusion obtained by LEP~\cite{LEPWorkingGroupforHiggsbosonsearches:2003ing}, where the $\sim 2\,\sigma$ excess around $95-98$~GeV can be seen. The dashed orange line indicates the improvements that can be expected at the ILC250 with an integrated luminosity of 2~ab$^{-1}$ according to the projection of \citere{Drechsel:2018mgd} (see also \citeres{Wang:2018fcw,Brudnowski:2024iiu,deBlas:2024bmz}). The blue and red points as well as the green star are as in the previous figures. It can be clearly observed that all parameter points within the preferred parameter region are well within the projected ILC250 sensitivity. Consequently, it is expected within the \megm\ model that the new Higgs boson at $\sim 95 \gev$ can be produced abundantly at the ILC250 (or other $e^+e^-$ colliders operating at $\sqrt{s} = 250$~GeV). These prospects are similar to the ones found in the GM analysis in \citere{Chen:2023bqr}.

%%%%%%%%%%%%%%%%%%%%%%%%%%% F I G U R E %%%%%%%%%%%%%%%%%%%%%%%%%%%%%%%%%%%%%%%%%%%%%%%%%%%%%%%%%%%%%%%%%%
%%%%%%%%%%%%%%%%%%%%%%%%%%% F I G U R E %%%%%%%%%%%%%%%%%%%%%%%%%%%%%%%%%%%%%%%%%%%%%%%%%%%%%%%%%%%%%%%%%%

\begin{figure}[ht!]
    \centering
    \includegraphics[width=0.75\textwidth]{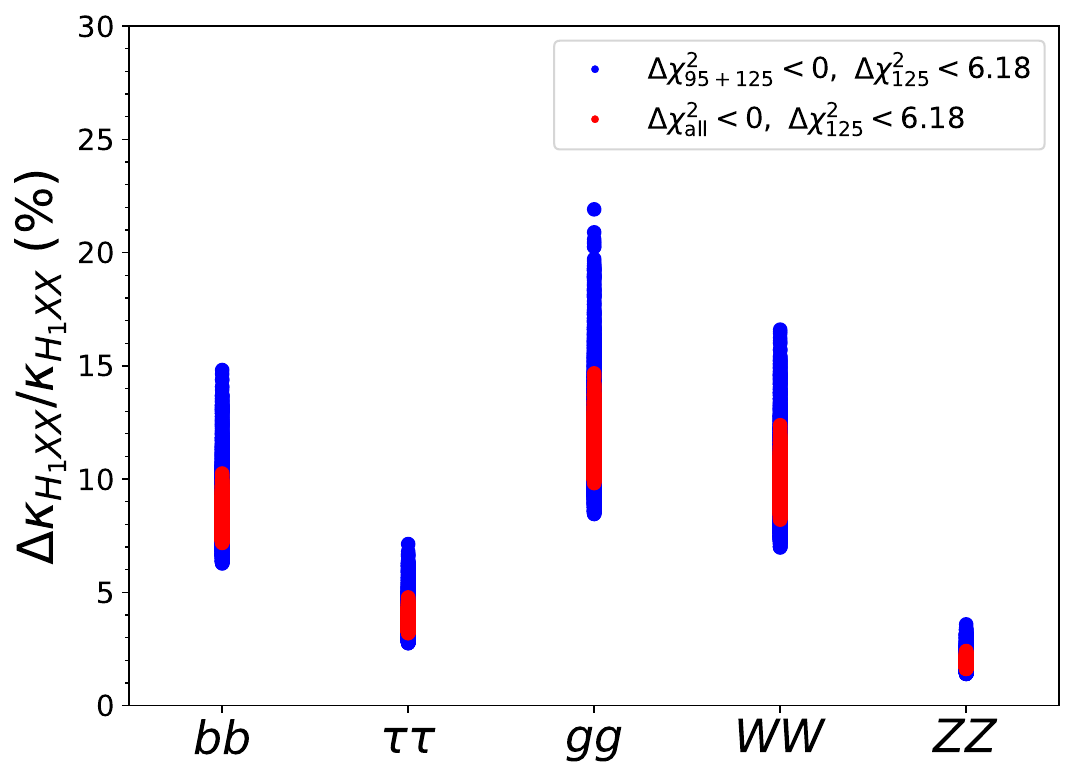}
    \caption{Sample distributions of $\Delta\kappa_{H_1XX}/\kappa_{H_1XX},~XX=b\bar{b},\tau\tau,gg,WW,ZZ$, expected from ILC250. The color coding is the same as in \protect\reffi{fig:vev-mu_H1}.
    }
    \label{fig:Delta_kappa}
\end{figure}

In \reffi{fig:Delta_kappa}, we analyze the prospects of $H_1$ coupling measurements at the ILC250. The evaluation of the anticipated precision of the coupling measurements is based on \citere{Heinemeyer:2021msz}. The evaluation has been performed for the (effective) couplings of $H_1$ to $b\bar b$, $\tau^+\tau^-$, $gg$, $WW$ and $ZZ$. While the first four rely on the decays of the Higgs boson to the respective final states, the $H_1ZZ$ coupling is obtained from the production of $H_1$ as radiated from a $Z$~boson. The latter channel yields the highest precision between $1-4\%$. A high accuracy is also expected for the coupling to $\tau$-leptons, ranging from $3-7\%$. The other three couplings are expected to be determined with an accuracy between $\sim 7\%$ and $\sim 22\%$. Coupling measurements at this level of precision will help to distinguish the \megm\ interpretation of the 95~GeV excesses from other model interpretations (see, \eg, \citeres{Heinemeyer:2021msz,Biekotter:2022jyr}, where prospective coupling precisions for the N2HDM and S2HDM have been evaluated).

%%%%%%%%%%%%%%%%%%%%%%%%%%% F I G U R E %%%%%%%%%%%%%%%%%%%%%%%%%%%%%%%%%%%%%%%%%%%%%%%%%%%%%%%%%%%%%%%%%%
%%%%%%%%%%%%%%%%%%%%%%%%%%% F I G U R E %%%%%%%%%%%%%%%%%%%%%%%%%%%%%%%%%%%%%%%%%%%%%%%%%%%%%%%%%%%%%%%%%%

A similar analysis could in principle be performed for the properties of $H_3$ at $152 \gev$. However, due to the fact that it is close to the kinematic limit of the ILC250 machine ($m_{H_3} + M_Z \approx 243 \gev \approx 250 \gev$) and to the reduced coupling of the $H_3$ to $Z$~bosons, only a marginal production cross section can be expected at this energy stage of the collider. Consequently, a higher center-of-mass energy would be necessary, e.g., the ILC with $\sqrt{s} = 500 \gev$ (ILC500) or LCF with 550~GeV. We leave an analysis of the properties of $H_3$ at $152 \gev$, corresponding to \reffis{fig:mH1_Lepton} and \ref{fig:Delta_kappa}, for future work.

%================================================================
\section{Conclusions}\label{sec:conclusions}

If both excesses in the LHC Higgs-boson searches at 95~and 152~GeV are confirmed by future data, the intriguing light scalar spectrum (including the 125~GeV Higgs boson) will immediately call for a natural interpretation within a well-defined model that is theoretically well motivated. We have demonstrated in the present paper that the meGM model is capable of offering a framework that describes well the excesses at 95~and 152~GeV, owing to the following features: the predictive spectrum of light scalar bosons, the enhancement in the ${\rm BR}(\phi\to\gamma\gamma)$ rates through the $H^{\pm\pm}$ loop, the asymmetric $\phi WW$ and $\phi ZZ$ couplings through custodial symmetry breaking, and the approximate preservation of $\rho=1$ at tree level by minimally extending the GM model configuration. As a result, the meGM model qualifies as a promising model for a further investigation on these topics.

As demonstrated in our numerical analysis, the meGM model yields a good fit to both the 95 and 152~GeV excesses: the fit to the data at about $95 \gev$ and $152 \gev$ as well as the overall fit is substantially improved in the meGM model compared to the case of the SM. At the same time, the meGM model in this scenario makes a clear prediction for a spectrum of additional CP-odd and charged scalars with masses below $\sim 170 \gev$. In particular, the presence of the light doubly-charged Higgs boson $H^{\pm\pm}$ is crucial for bringing the predicted di-photon rates into the observed ranges. While the treatment of the experimental situation regarding the excess at about 152~GeV has been done in a simplified way in this paper, dedicated experimental analyses in the future have potential for substantial improvements.

In addition to further tests of the light neutral excesses via direct searches, we have also pointed out the potential of searches for additional charged scalars predicted by the meGM model in future runs of the LHC for probing this very predictive scenario. As a result of the mild custodial symmetry breaking in the meGM model, the predicted production cross sections of $H_2^\pm$ and $H^{\pm\pm}$ are both increased compared to their counterpart processes in the GM model, thus providing a better chance of discovery. Finally, we have demonstrated that precision coupling measurements of the Higgs bosons at $95$ and $125 \gev$ at the HL-LHC or a future $e^+e^-$ collider (where we have used the ILC250/LCF250 as a concrete example) can yield important information to further scrutinize the meGM interpretation of the observed excesses.

%================================================================
% Acknowledgements
%================================================================
\section*{Acknowledgments}

We thank A.~Crivellin and B.~Mellado for interesting discussions.
T.K.C. is supported by the Ministry of Education, Taiwan, under the Government Scholarship to Study Abroad.
C.W.C. is supported in part by the National Science and Technology Council under Grant Nos.~NSTC-111-2112-M-002-018-MY3 and NSTC114-2112-M-002-020-MY3.
The work of S.H.\ has received financial support from the
grant PID2019-110058GB-C21 funded by
MCIN/AEI/10.13039/501100011033 and by ``ERDF A way of making Europe'', and in part by by the grant IFT Centro de Excelencia Severo Ochoa CEX2020-001007-S funded by MCIN/AEI/10.13039/501100011033.
S.H.\ also acknowledges support from Grant PID2022-142545NB-C21 funded by MCIN/AEI/10.13039/501100011033/ FEDER, UE.
G.W.\ acknowledges the support of the Deutsche Forschungsgemeinschaft (DFG, German Research Association) under Germany's Excellence Strategy-EXC 2121 ``Quantum Universe''-390833306. This work has also been funded by the Deutsche Forschungsgemeinschaft (DFG, German Research Foundation) -- 491245950.

%================================================================
% Appendix
%================================================================
\begin{appendix}
%=================================================================================================
\section{Mass Formulas}\label{sec:mass}

We provide here the explicit formulas for masses or mass matrices of the physical Higgs bosons based on the general Higgs potential given in Eq.~\eqref{eq:pot_gen} without imposing any assumptions.

First, the squared mass of the doubly-charged Higgs bosons $\chi^{\pm\pm}$ is given by 
\begin{align}
m_{\chi^{\pm\pm}}^2 &= -2 \rho_2 v_\chi^2 - \frac{\sigma_2}{2}v_\phi^2 -\sqrt{2} \mu_{\chi \xi}  v_\xi-\frac{v_\phi^2}{4 }\left(2\frac{\Re\mu_{\phi \chi}}{v_\chi} +\sqrt{2}\Re\sigma_4 \frac{v_\xi}{v_\chi} \right) ~.
\end{align}

Denoted by $M^2_\pm$ and $M^2_0$, respectively, the Hermitian mass-squared matrices for the singly-charged and neutral Higgs bosons in the basis of $(\tilde{H}_1^\pm,\tilde{H}_2^\pm)$ and $(\tilde{h},\tilde{H}_1,\tilde{H}_2,\tilde{H}_3)$ (see \refeq{eq:masseigen1} for the definition of these fields) 
have the matrix elements given by
\begin{align}
\begin{split}
(M^2_\pm)_{11} 
=& 
-\frac{v^2}{4(v_\xi^2+v_\chi^2)}\left[\sigma_2   v_\chi^2 + \sqrt{2}\Re \sigma_4 v_\chi v_\xi -\sqrt{2}\mu_{\phi \xi}v_\xi + 2\Re\mu_{\phi \chi}v_\chi \right]~, \\
(M^2_\pm)_{22} 
=&
\frac{v_\xi^2+v_\chi^2}{2} \left(2 \rho_5  - \sqrt{2} \frac{\mu_{\chi\xi}}{v_\xi} \right) 
\\
&
-\frac{v_\phi^2 }{4(v_\xi^2+v_\chi^2)}\left[ v_\xi^2\left(\sigma_2 + 2\frac{\Re\mu_{\phi \chi}}{v_\chi}\right)   -\sqrt{2}v_\chi^2\frac{\mu_{\phi\xi}}{v_\xi}  
+ \sqrt{2}\Re \sigma_4 v_\xi v_\chi\left(2 + \frac{v_\xi^2}{v_\chi^2} + 2\frac{v_\chi^2}{v_\xi^2}  \right)  \right]~, \\
(M^2_\pm)_{12} 
=&
-\frac{v_\phi v}{4 (v_\xi^2+v_\chi^2)}v_\chi v_\xi \left[\sigma_2 + 2\frac{\Re\mu_{\phi \chi}}{v_\chi} + 
  \sqrt{2} \frac{\mu_{\phi\xi}}{v_\xi}  -\sqrt{2} \frac{v_\chi}{v_\xi}\Re \sigma_4 \right]  + i \frac{v_\phi v}{2\sqrt{2}}\Im \sigma_4 ~, 
\end{split}
\end{align}
and 
\begin{align}
\begin{split}
(M_0^2)_{11} & = 2v_\phi^2\lambda ~, \\
(M_0^2)_{22} & = 4v_\chi^2(\rho_1 + \rho_2) - \frac{v_\phi^2}{4}\left(\sqrt{2}\frac{v_\xi}{v_\chi}\Re \sigma_4 + \frac{1}{2}\frac{\Re\mu_{\phi\chi}}{v_\chi} \right) ~, \\
(M_0^2)_{33} & = 8v_\xi^2\rho_3 - \frac{v_\chi^2}{2\sqrt{2}}\frac{\mu_{\chi\xi}}{v_\xi} - v_\phi^2 \frac{v_\chi}{\sqrt{2} v_\xi}\Re\sigma_4 + \frac{v_\phi^2}{2\sqrt{2}}\frac{\mu_{\phi\xi}}{v_\xi} ~, \\
(M_0^2)_{44} & = -\frac{v_\phi^2 + 8v_\chi^2}{4}\left(2\frac{\Re\mu_{\phi\chi}}{v_\chi} + \frac{\sqrt{2}v_\xi}{v_\chi} \Re \sigma_4 \right) ~, \\
(M_0^2)_{12} & = \sqrt{2} v_\phi v_\chi \left(\sigma_1 + \sigma_2\right) + v_\phi v_\xi\Re \sigma_4 + \sqrt{2}v_\phi \Re\mu_{\phi\chi} ~,  \\
(M_0^2)_{13} & = 2v_\phi v_\xi \sigma_3+\sqrt{2}v_\phi v_\chi \Re\sigma_4 -\frac{v_\phi \mu_{\phi\xi}}{\sqrt{2}} ~, \\
(M_0^2)_{14} & = 0 ~, \\
(M_0^2)_{23} & = 2\sqrt{2}v_\chi v_\xi \rho_4 + \frac{v_\phi^2}{2}\Re\sigma_4 + v_\chi \mu_{\chi\xi} ~, \\
(M_0^2)_{24} & = 0 ~, \\
(M_0^2)_{34} & = -\frac{v_\phi}{2}\sqrt{v_\phi^2 + 8v_\chi^2}\Im \sigma_4 ~. 
\end{split}
\end{align}

%================================================================
\end{appendix}


%================================================================
\end{document}